\documentclass[prd,aps,preprintnumbers,superscriptaddress,showpacs,eqsecnum,nofootinbib,twocolumn,showkeys]{revtex4-1}
\usepackage{blkarray,graphicx,dcolumn,bm,hyperref,physics,here,mathrsfs,cases,empheq,latexsym,amsfonts,color}

\renewcommand{\curl}{\operatorname{curl}}
\renewcommand{\div}{\operatorname{div}}
\newcommand{\Lie}{\pounds}

\begin{document}

\allowdisplaybreaks[1]
\preprint{YITP-23-113, IPMU23-0032}
\title{On spin optics for gravitational waves lensed by a rotating object}
\author{Kei-ichiro Kubota}
    \email{keiichiro.kubota@yukawa.ktoto-u.ac.jp}
    \affiliation{Center for Gravitational Physics and Quantum Information, Yukawa Institute for Theoretical Physics, Kyoto University, 606-8502, Kyoto, Japan.}
\author{Shun Arai}
    \affiliation{
    Kobayashi-Maskawa Institute, Nagoya University, Nagoya 464-8602, Japan}
\author{Shinji Mukohyama}
    \affiliation{Center for Gravitational Physics and Quantum Information, Yukawa Institute for Theoretical Physics, Kyoto University, 606-8502, Kyoto, Japan.}
    \affiliation{Kavli Institute for the Physics and Mathematics of the Universe (WPI), The University of Tokyo Institutes for Advanced Study, The University of Tokyo, Kashiwa, Chiba, 277-8583, Japan.}
\date{\today}
\begin{abstract}
    We study gravitational lensing of gravitational waves taking into account the spin of gravitational waves coupled with a dragged spacetime made by a rotating object. We decompose the phase of gravitational waves into helicity-dependent and independent components with spin optics, analyzing waves whose wavelengths are shorter than the curvature radius of a lens object. 
    We analytically confirm that the trajectory of gravitational waves splits depending on the helicity, generating additional time delay and elliptical polarization onto the helicity-independent part. 
    We exemplify monochromatic gravitational waves lensed by a Kerr black hole and derive the analytical expressions of corrections in phase and magnification.
    The corrections are enhanced for longer wavelengths, potentially providing a novel probe of rotational properties of lens objects in low-frequency gravitational-wave observations in the future.  
\end{abstract}
\maketitle

\section{\label{sec:intro}Introduction}
Gravitational waves carry rich information to probe the Universe and fundamental physics:\ particle physics, nuclear physics, and tests for gravity theories. The network of the ground-based interferometry of
LIGO-Virgo-KAGRA 
has detected gravitational waves from nearly a hundred of
steller-mass binary mergers~\cite{LIGOScientific:2021djp}. Such data make it possible to examine the physical models of neutron stars and black holes at the extreme blink of coalescence. Meanwhile, the planned space-based detectors e.g. Laser Interferometer Space Antenna (LISA)~\cite{LISA:2017pwj} and the DECi-hertz Interferometer Gravitational-wave Observatory 
 (DECIGO) \cite{Kawamura:2006up} (see the current status reported in \cite{Kawamura:2020pcg}) explore gravitational waves in low-frequency band, spanning 1mHz - 1Hz and 0.1Hz - 10Hz respectively, in order to probe binary evolution, background radiation, and gravitational-waves propagation. In this paper, we consider wave propagation in order to understand the wave nature of gravitational waves.

One of the standard ways to describe gravitational-wave propagation is to employ the geometrical optics for the fiducial Einstein's general relativity, as the wavelength is well shorter than the size of a lens object in most of the situations during propagation\footnote{The cases where the short-wavelength approximation is no longer accurate, i.e., wave optics, has been studied in e.g. Ref.~\cite{Nakamura:1999uwi, Takahashi:2003ix}. These effects are beyond our scope in this paper.}. In geometrical optics, Einstein's general relativity 
predicts that (1) gravitational waves propagate at the speed of light in a vacuum, (2) gravitational waves are lensed in the same way as electromagnetic waves are, and (3) gravitational waves possess the two tensorial polarization modes. These predictions are able to be examined with observations. In fact, the arrival-time difference between gravitational waves and electromagnetic waves of the binary neutron star merger associated with a gamma-ray burst: GW170817/GRB170817A~\cite{LIGOScientific:2017vwq} was measured, confirming that the gravitational waves propagate at the speed of light with $10^{-15}$ precision. This gives strong constraints on alternative theories of gravity that potentially explain cosmic acceleration~\cite{Creminelli:2017sry,Ezquiaga:2017ekz,Sakstein:2017xjx,Baker:2017hug,Arai:2017hxj,Amendola:2017orw}. 
The polarization of gravitational waves has been examined by the LIGO-Virgo-KAGRA network of interferometry, constraining an upper bound on non-standard polarization modes~\cite{LIGOScientific:2017ycc,LIGOScientific:2018dkp,Takeda:2020tjj,LIGOScientific:2020tif,Hagihara:2019ihn,Pang:2020pfz,Takeda:2021hgo}.

We aim to explore further detailed physics of gravitational-wave propagation in Einstein's general relativity, focusing on the coupling between the spin of gravitational waves and rotational components of the background spacetime on which waves propagate. Since this coupling cannot be captured by geometrical optics, we develop a way to derive the analytic solution in the regime where the linear perturbation is valid. Provided the linear perturbation works, one can always decompose a gravitational wave into any basis, such as a set of monochromatic waves or a set of wave packets, and then study the propagation of each element of the chosen basis. One can then compute the waveform at the position of the detector by taking an appropriate linear combination and the result should of course be independent of the choice of basis. In the present paper we study monochromatic waves, for which the propagation in the regime of validity of linear perturbation is fully characterized by phases and magnifications once the propagation of polarization tensors is specified.

There are various approaches to describe the propagation around a rotating object taking the spin into account.
One of the approaches is based on the gravitational Faraday rotation~\cite{Frolov:2011mh,Yoo:2012vv,Dolan:2017zgu,Dolan:2018ydp,Frolov:2012zn}.
This approach treats the gravitational Faraday rotation angle in the phase so that the equation of motion appropriately determines the evolution of polarization.
The authors of the papers called this approach ``spin optics''. We follow this jargon throughout this paper.
Other approaches are the Souriau-Saturunini equations~\cite{AIHPA_1974__20_4_315_0,saturnini:tel-01344863,Duval:2018hzh,Duval:2016hxo}, which are similar form to the Mathisson-Papapetrou-Dixon equations~\cite{2010GReGr..42.1011M,1951RSPSA.209..248P,1964NCim...34..317D,Dixon:2015vxa} for massless particles, and the Berry phase approach~\cite{Yamamoto:2017gla,Oancea:2020khc,Oancea:2022szu,Andersson:2023bvw,Andersson:2020gsj} (see Ref.~\cite{Oancea:2019pgm,Harte:2022dpo} for the relation between these approaches).
Note that the propagation of monochromatic gravitational waves is considered in the spin-optics approach, whereas localized wave packets are considered in the Souriau-Saturunini equation and Berry phase approach. 
As already mentioned above, gravitational waves in the regime of validity of linear perturbation can be decomposed into either a set of monochromatic waves or a set of wave packets, the propagation of each element of the chosen basis can be studied separately and one can finally take a linear combination to compute the waveform at the position of the detector.

The effect of spin tends to be more enhanced for longer-wavelength gravitational waves as pointed out analytically in ~\cite{Yamamoto:2017gla} and numerically~\cite{Oancea:2020khc,Oancea:2022szu,Andersson:2023bvw,Oancea:2023hgu,Dahal:2023ncl}, although these papers consider the propagation of wave packet different from the monochromatic waves we focus on.
Furthermore, a study that investigates the scattering of gravitational waves by a Kerr black hole using black hole perturbation theory has shown that the spin effect on the scattering amplitude becomes more pronounced for longer wavelengths~\cite{Dolan:2008kf,Leite:2017zyb,Leite:2018mon}.
Thus, the spin effect tends to be important for long-wavelength gravitational waves in addition to the wave effects.
However, despite their focus on the wave effect of long-wavelength gravitational waves, several studies aiming to determine the mass distribution of lens objects by using the wavy nature of gravitational waves~\cite{Takahashi:2003ix,Tambalo:2022wlm,Leung:2023lmq} have neglected the evolution of the polarization tensor, i.e., neglected the effect of spin.
In other words, the gravitational waves are treated as scalar waves following a null geodesic.
The treatment may not be correct and hence its validity needs to be investigated.

Ideally, one would like to completely incorporate both wave and spin effects for gravitational waves whose wavelength is not necessarily shorter than the radius of curvature of a lens object, whilst it is a challenge. In this paper, as a stepping stone towards the ideal calculation, we propose a method to calculate the first-order correction to the phase difference and magnification taking into account the spin effect induced by dragged components of spacetime for monochromatic gravitational waves with a wavelength shorter than the curvature radius of the lens object.
This effect has not been taken into account in a previous study e.g. Ref.~\cite{Baraldo:1999ny}, whereas it has recently been addressed in \cite{Li:2022izh} by using the Walker-Penrose theorem to compute the gravitational Faraday rotation. We apply our method to gravitational waves lensed by a Kerr black hole in spin optics and demonstrate its practical implementation for future observations.
Our results reveal two new points.
One is that there is an arrival time difference between left- and right-handed gravitational waves.
The other is that linear polarized gravitational waves lensed by a rotating lens object generally tend to be elliptically polarized for longer wavelength gravitational waves, supporting the results of the study about the scattering by a Kerr black hole\cite{Dolan:2008kf}.

The rest of this paper is organized as follows.
In \S~\ref{sec:formalism}, we prepare the way to analytically calculate the phase and magnification of left- and right-handed gravitational waves. We demonstrate the application of our method to the gravitational wave lensed by a Kerr black hole in \S~\ref{sec:examplekerr}.
Finally, we conclude the paper in \S~\ref{sec:conclusion}.

Let us introduce the notation that we use throughout the paper. The indices of tensors $a,b,c,d,e,f,g,h$ run over $0$ to $3$, whereas spatial indices $i,j,k,l$ run over $1$ to $3$.
The round and square brackets of indices denote symmetrization and antisymmetrization, respectively, that is, $T_{(ab)}:=(T_{ab}+T_{ba})/2!$ and $T_{[ab]}:=(T_{ab}-T_{ba})/2!$.
$\eta_{abcd}$ and $\epsilon_{abc}$ denote a 4-dimensional and spatially completely antisymmetric tensor, respectively.
The variables with a bar denote the helicity-independent part of the variables.
The bold math symbols denote spatial vectors and operators.
We use the unit $c=G=1$.

\section{\label{sec:formalism}Formalism}

We aim to derive the arrival-time difference and the elliptical polarization of gravitational waves induced by the coupling between spin and the rotational component of the background spacetime.  
We define the metric as
\begin{equation}
    {\mathrm d}s^2 = g^{(\text{phy})}_{ab}{\mathrm d}x^a{\mathrm d}x^b\,,
\end{equation}
where $g^{(\text{phy})}_{ab}$ is decomposed into the background and the perturbation as

\begin{equation}
    g^\text{(phy)}_{ab}=g_{ab}+h_{ab}.
\end{equation}
In order to extract the physical degrees of freedom that correspond to the gravitational waves, we impose the transverse-traceless gauge i.e. $h^a{}_a=0=\nabla^a h_{ab}$. Then the Einstein equation $G_{ab}[{g}^\text{(phy)}_{cd}] = 0$ is linearized and obtain the wave equation as
\begin{align}
    \Box h_{ab}+2R_{acbd}h^{cd}=0,
    \label{eq:EoM}
\end{align}
where $\Box:=g^{ab}\nabla_a\nabla_b$. This equation is what we solve in the whole paper.

We assume that the background spacetime on which gravitational waves propagate is approximately stationary, considering a situation where the rotational motion of a lens object distorts the background much slower than the time variation of gravitational waves. The stationary conditions simplify the discussion without the loss of critical properties of spin optics i.e. the helicity-dependent split of gravitational-wave trajectory.

Let the background spacetime be a stationary spacetime throughout this paper.
The stationary spacetime means that there exists a timelike Killing vector $\xi_{(t)}^a$ parameterized by $t$, i.e.,
\begin{align}
    &\exists \xi_{(t)}^a=(\partial_t)^a \text{ s.t. } \pounds_{\xi_{(t)}} g_{ab} = 0 \text{ and }\xi_{(t)}^a \xi^{(t)}_a <0\nonumber\\
    \Leftrightarrow & \exists \text{time coordinate }t \text{ s.t. } \partial_t g_{ab} = 0.
\end{align}
The timelike hypersurface with constant $t$, $\Sigma_t := \mathcal{M}/\mathcal{G}$ is the orbit space associated with $\xi_{(t)}^a$, where $\mathcal{M}$ is the background spacetime manifold and $\mathcal{G}$ is the isometry group of transformations generated by $\xi_{(t)}^a$. We express the metric of the background spacetime as
\begin{align}
    g_{ab} \mathrm{d}x^a \mathrm{d}x^b = g_{tt}(\mathrm{d}t - g_i \mathrm{d}x^i)^2 + \gamma_{ij}\mathrm{d}x^i \mathrm{d}x^j ,
\end{align}
where $g_{tt}$, $g_i$, and $\gamma_{ij}$ are the functions of the spatial coordinate $x^i$ and independent of the time coordinate $t$.
The components of the inverse of the metric are $g^{tt}=1/(g_{tt})+g_ig^i$, $g^{ti}=g^i$, and $g^{ij}=\gamma^{ij}$.
The determinant is $\det(g_{ab})=g_{tt}\det(\gamma_{ij})$.
We introduce a normalized Killing vector $u^a$ as
\begin{align}
    u^a:=\frac{\xi_{(t)}^a}{\sqrt{-g_{tt}}}.
    \label{eq:timelikenormalvector}
\end{align}
By definition, $u^au_a=-1$.
We define the induced metric $\gamma_{ab}$ by using $u^a$ as
\begin{align}
    \gamma_{ab}:=g_{ab}+u_a u_b.
\end{align}
We also define the covariant derivative $\mathrm{D}_a$ associated with $\gamma_{ab}$ as
\begin{align}
    &\mathrm{D}_a S^{b_1 b_2 \cdots b_k}_{\ \ \ \ c_1 c_2  \cdots c_l} \nonumber \\
    &:= \gamma^{\ c}_{a}\gamma_{c_1}^{\ e_1}\gamma_{c_2}^{\ e_2}\cdots\gamma_{c_l}^{\ e_l}\gamma_{\ d_1}^{b_1}\cdots\gamma_{\ d_k}^{b_k}\nabla_c S^{d_1 d_2 \cdots d_k}_{\ \ \ \ e_1 e_2  \cdots e_l},
\end{align}
where $S^{b_1 b_2 \cdots b_k}_{\ \ \ \ c_1 c_2  \cdots c_l}$ is a spatial tensor, and $\nabla_a $ is the covariant derivative associated with $g_{ab}$.

We employ the spin optics and the diffraction formula allowed to analytically calculate the phase of gravitational waves along the trajectory. Spin optics is available in situations where the wavelength is shorter than the curvature radius of the lens object, whereas not capturing all the wave effects.
However, it is interesting to consider such situations because the time delay and the elliptical polarization generated from the rotational dragging by a lens object are analytically understood.

\subsection{\label{sec:spinoptics}spin optics}
As conventionally known in the context of gravitational Faraday rotation ~\cite{Ishihara:1987dv,Nouri-Zonoz:1999jls,Chakraborty:2021bsb,Tamburini:2021jok,Dolan:2017zgu,Dolan:2018ydp}, the polarization vectors are revolved through the coupling between the polarization vectors and the rotational dragging of the background spacetime, generating a rotational angle.
Spin optics treats the rotational angle as a helicity-dependent additional phase shift.
Following the previous studies~\cite{Frolov:2011mh,Frolov:2012zn,Yoo:2012vv,Frolov:2020uhn}, we employ the polarization base tensor extended by the Fermi-Walker parallel transport.

In this section, we define the basis as the circular polarization vectors, the Fermi-Walker parallel transport, and the modified dispersion relation in \S~\ref{sec:circularpolarization}, and \S~\ref{sec:dispersionrelation}, respectively. In \S~\ref{sec:dispersionrelation}, we analytically solve the dispersion relation for the modified phase.
In \S~\ref{sec:GFRA}, we supply an alternative explanation of the spin effect in the language of the gravito-electromagnetism.

\subsubsection{\label{sec:circularpolarization}Circular polarization basis}
\begin{figure}[t]
    \includegraphics[width=1.0\columnwidth]{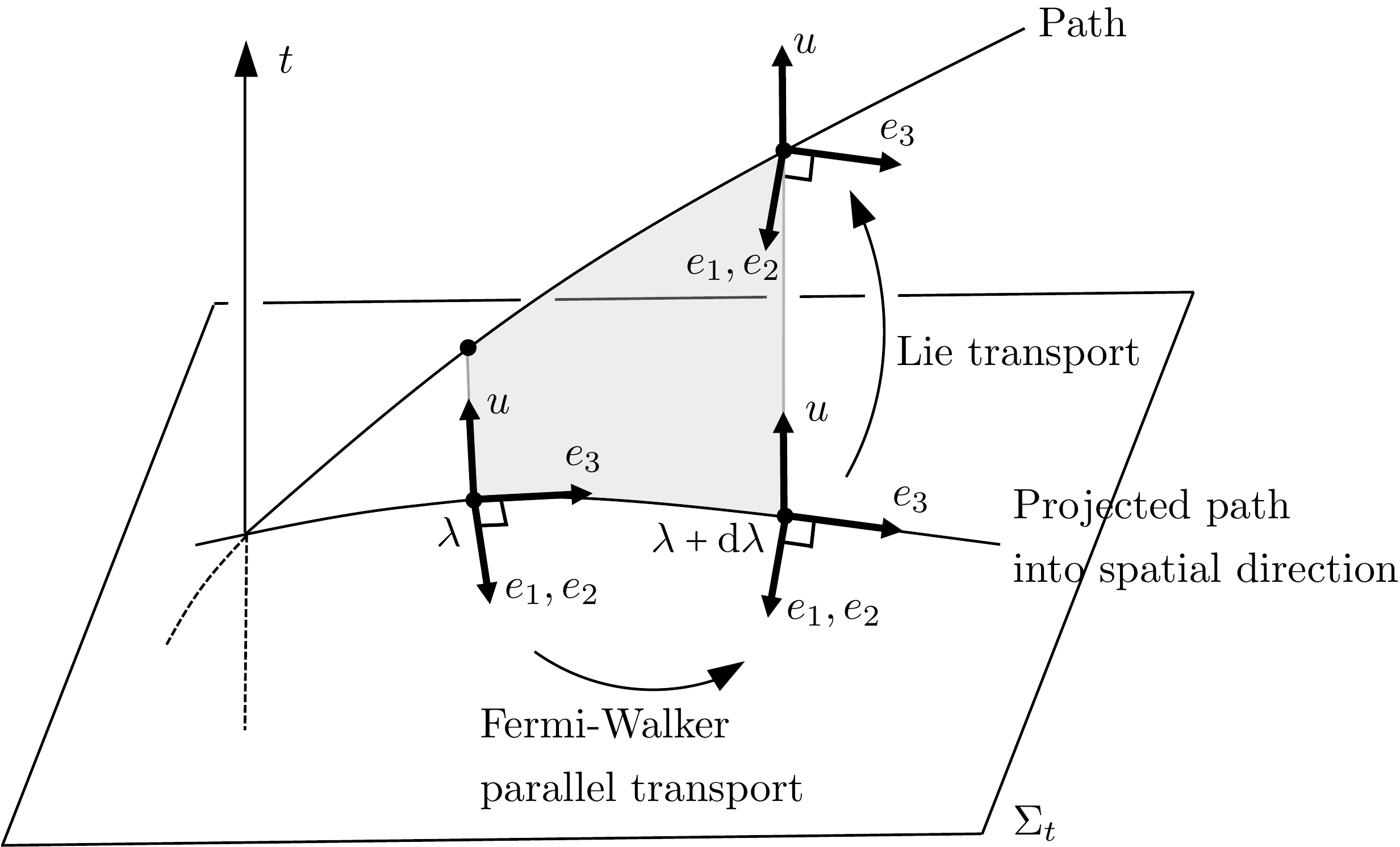}
    \caption{\label{fig:base_transport}Transportation of the orthonormal base vectors.}
\end{figure}
We introduce a tetrad $\{u^a,e_1{}^a,e_2{}^a,e_3{}^a\}$ associated with the trajectory of the gravitational waves (see Fig.~\ref{fig:base_transport}) which satisfy the unit orthogonal conditions,
\begin{align}
    e_i{}^a e_j{}_a &= \delta_{ij},& e_i{}^a u_a = 0.
\end{align}

As depicted in Fig.~\ref{fig:base_transport}, we extend the base vectors ${e_i}^a$ in the time direction by the Lie transport along the integral curve of $\xi^a_{(t)}$ as
\begin{align}
    \Lie_{\xi_{(t)}}{e_i}^a = 0.
    \label{eq:Lietransportofbasevectors}
\end{align}
We extend the base tensors in the spatial direction by the spatial Fermi-Walker parallel transport.
The projected path of the non-geodesic path generally is not spatial geodesic, i.e. $e_3^a\mathrm{D}_a {e_3}^b \neq 0$, then the base tensor transported by the spatial standard parallel transport does not point in the tangent direction of the spatial path.
To get around this problem, we extend the base vectors ${e_i}^a$ in the space direction by spatial Fermi-Walker parallel transport along the spatial path as
\begin{align}
    \mathrm{D}_{e_3}^\text{(FW)}{e_i}^a=0,
     \label{eq:FWtransportofbasevectors}
\end{align}
where $\mathrm{D}_{e_3}^\text{(FW)}$ denotes the spatial Fermi-Walker derivative~\footnote{Note that the sign of $\mathcal{F}_{ab}$ oppositely flips between space-like and time-like Fermi-Walker parallel transport so that making inner products of Fermi-Walker parallel transport vectors invariant.} defined by
\begin{align}
    \mathrm{D}_{e_3}^\text{(FW)}{e_i}^a:=&{e_3}^c\mathrm{D}_c {e_i}^a+\mathcal{F}^{a}{}_c {e_i}^c, \nonumber \\
    \mathcal{F}_{ab}:=&{e_3}_a a_b - {e_3}_b a_a,
\end{align}
and $a^a:={e_3}^b \mathrm{D}_b {e_3}^a$ is the acceleration for ${e_3}^a$.
The base tensor transported by the spatial Fermi-Walker parallel transport is tangent to the spatial path (see App.~\ref{sec:FermiWalkertransport}).

Next, we define the circular polarization base vector using the tetrad.
The circular base vector $m^a$ is defined by
\begin{align}
    m^{a}:=\frac{1}{\sqrt{2}}({e_1}^{a}+i {e_2}^{a}).
\end{align}
By definition, the circular base vector satisfies
\begin{align}
    m_{a}m^{*a}=&1,&m_{a}m^{a}=&m_a {e_3}^a =m_a u^a =0,
    \label{eq:circularvectorcondition}
\end{align}
and
\begin{align}
    m^{a}m^{*b}=&-i e_1^{[a}e_2^{b]}+\frac{1}{2}(e^a_1e^b_1+e^a_2 e^b_2).
    \label{eq:circularandbasevector}
\end{align}
Here, the asterisk denotes the complex conjugate.
The condition $m^a m_a = 0$ yields
\begin{align}
    m_{a}\nabla_c m^{a}=0.
    \label{eq:derivativeofbasevectorcondition}
\end{align}
The circular polarization tensor is transported in the same way as the tetrad base tensors,
\begin{align}
    \Lie_{\xi_{(t)}}m^{a} =& 0 ,\label{eq:Lietransportcircular}\\
    \mathrm{D}_{e_3}^\text{(FW)}m^{a}=& 0.\label{eq:FWtransportcircular}
\end{align}
Note that the above extension of the base vectors is valid for stationary spacetime.
The extensions for the arbitrary asymptotically-flat spacetime are discussed in Ref.~\cite{Shoom:2020zhr}.

\subsubsection{\label{sec:dispersionrelation}Dispersion relation}
We introduce the dispersion relation in spin optics which incorporates the effect of spin on the trajectory.
We compute the dispersion relation only for the right-handed gravitational waves for the brevity of presentation.
Performing complex conjugate in the following discussion gives us a discussion for a left-handed gravitational waves.

Considering the path away from the lens, we neglect the second term in the left-hand side of Eq.~\eqref{eq:EoM} throughout this paper. Provided that $|\nabla \mathcal{A}/\mathcal{A}|$, $|\nabla m_{ab}/m_{ab}|\ll |\nabla S/S|$, the right-handed metric perturbation can be decomposed into the amplitude $\mathcal{A}_R$, the polarization base tensor $m_{ab}:=m_a m_b$, and the phase $S_R$ as
\begin{align}
    h_{\text{R}ab}=\mathcal{A}_\text{R}m_{ab}e^{iS_\text{R}}.
\end{align}
The indices ``R'' and ``L'' denote right- and left-handed, respectively.
We use these indices only for polarization-specific discussion and omit them when the polarization is not specific.

Under the geometrical optics assumption, the equation of motion~\eqref{eq:EoM} becomes
\begin{align}
    &\Bigl(\mathcal{A}_\text{R} m_{ab}\nabla^c S_\text{R} \nabla^c S_\text{R} -2i\mathcal{A}_\text{R}\nabla^c S_\text{R} \nabla_c m_{ab} \nonumber \\ 
    &-i\mathcal{A}_\text{R} m_{ab}\nabla^c \nabla_c S_\text{R}  - 2im_{ab}\nabla^c S_\text{R} \nabla_c \mathcal{A}_\text{R}\Bigr)e^{iS_\text{R}} \nonumber \\ 
    &+ \mathcal{O}\left((\nabla S_\text{R})^0\right)=0
    \label{eq:eomseries}
\end{align}
In the standard geometrical optics, one collects the term with $O((\nabla S_\text{R})^2)$ and hence obtains the standard dispersion relation $\nabla_a S_\text{R} \nabla^a S_\text{R} = 0$ from the leading order of Eq.~\eqref{eq:eomseries}.
In spin optics, on the other hand, the next-to-leading order term $-2i\mathcal{A}_\text{R}\nabla^c S_\text{R} \nabla_c m_{ab}$ is regarded as the correction to the dispersion relation. This term physically describes the modulation of polarization tensor along the trajectory, namely the gravitational Faraday rotation. Note that this term flips its signature on the left-handed mode.
Contracting Eq.~\eqref{eq:eomseries} with $m^{*ab}$, the dispersion relation and the Hamiltonian in spin optics are given by
\begin{align}
    H:=\frac{1}{2}g^{ab}\left(\nabla_a S - \sigma\mathcal{B}_a\right)\left(\nabla_b S -\sigma\mathcal{B}_b\right) \approx 0.
    \label{eq:spinopticsdispersionrelation}
\end{align}
where the helicity $\sigma$ take $+2$ for the right- and $-2$ for the left-handed gravitational waves.
``$\approx$'' denotes the equality up to the first order of $\mathcal{B}$ where  $\mathcal{B}_a$ is defined by
\begin{align}
    \mathcal{B}_a:=i m^{*b}\nabla_a m_b.
\end{align}
The key is to use this dispersion relation instead of the standard dispersion relation for massless particles in order to take into account the spin effect.
The next leading order of Eq.~\eqref{eq:eomseries} yields the conservation of the energy of gravitational waves as in the same as standard geometrical optics approximation~\cite{Yoo:2012vv}.

We define the wave vector and velocity as
\begin{align}
    k_a:=&\nabla_a S,\\
   \dot{x}^a:=&\frac{\partial H}{\partial k_a}= k^a -\sigma\mathcal{B}^a.
   \label{eq:xdotandk}
\end{align}
Here, we take the $e_3{}^a$ as the direction of the spatial part of the velocity.
The explicit expression of $e_3{}^a$ is given by
\begin{align}
    {e_3}^a :=\frac{\sqrt{-g_{tt}}}{-\xi_{(t)}^b \dot{x}_b}\dot{x}^a - u^a.
    \label{eq:spatialnormalvector}
\end{align}
The coefficient of $\dot{x}^a$ is determined so that the approximate null condition Eq.~\eqref{eq:spinopticsdispersionrelation} is satisfied.

Because of stationarity, we impose
\begin{align}
    \pounds_{\xi_{(t)}}\dot{x}^a =& \partial_t \dot{x}^a=0\,,
    \label{eq:Lieofxdot}
\end{align}
indicating that the phase velocity stays constant in time. 
Since $\mathcal{B}_a$ consists of the circular base vector which is Lie transported along the Killing vector $\xi^a_{(t)}$, $\mathcal{B}_a$ is also Lie transported, i.e., $\Lie_{\xi_{(t)}}\mathcal{B}^a=0$.
Indeed, we immediately show $\Lie_{\xi_{(t)}}\mathcal{B}^a=0$, because the Lie derivative along any Killing vector commutes with any covariant derivative, i.e., $[\Lie_\xi,\nabla]=0$, (See App.~\ref{sec:Commutation}).
It means
\begin{align}
    \Lie_{\xi_{(t)}} \mathcal{B}^a = \partial_t \mathcal{B}^a =0.
\end{align}
This equation and Eqs.~\eqref{eq:Lieofxdot}~\eqref{eq:xdotandk} lead to 
\begin{align}
    \Lie_{\xi_{(t)}}k^a=\partial_t k^a=0.
    \label{eq:timederivativeofpartialS}
\end{align}
We define the frequency of gravitational waves $\omega$ as
\begin{align}
    \omega := - \xi_{(t)}^a k_a.
\end{align}
Using Eq. \eqref{eq:timederivativeofpartialS} and $\partial_t \xi_{(t)}^a =0$, the derivative of the frequency is 
\begin{align}
    \partial_a \omega =-\xi_{(t)}^b\partial_a k_b - k_b\partial_a \xi_{(t)}^b =0.
\end{align}
Thus we obtain
\begin{align}
    \omega = \text{constant}.
\end{align}
This is the consequence of the temporal isometry of the background spacetime.
We can separate the phase into the time-dependent part $-\omega t$ and space-dependent part $\mathcal{S}(x^i)$ as
\begin{align}
    S(t,x^i) = -\omega t + \mathcal{S}(x^i).
\end{align}

To analytically solve the dispersion relation~\eqref{eq:spinopticsdispersionrelation},  we additionally assume that the frequency is larger than $|\mathcal{B}_a|$, i.e., $|\mathcal{B}_a|\ll|k_a|$ on the path.
Let set $S=\bar{S}+\sigma \chi$ where $\bar{S}$ and $\sigma \chi$ are the zeroth and first-order solutions of $\mathcal{B}$, respectively, where
$\bar{S}$ is the helicity-independent part of the $S$.
Note that the helicity-dependent part $\sigma \chi$ is contained in the spatial part of the phase $\mathcal{S}$.

The zeroth order of $\mathcal{B}$ of Eq.~\eqref{eq:spinopticsdispersionrelation} yields
\begin{align}
    \nabla_a \bar{S} \nabla^a \bar{S} = \bar{k}^a \nabla_a \bar{S} \approx 0,
    \label{eq:zerothorderofdisperaisionrelation}
\end{align}
where $\bar{k}_a:=\nabla_a \bar{S}$.
The covariant derivative of this equation yields the geodesic equation,
\begin{align}
    \bar{k}^b\nabla_b \bar{k}^a\approx 0.
    \label{eq:geodesic}
\end{align}
Eq.~\eqref{eq:zerothorderofdisperaisionrelation} means $\bar{S}$ is constant along the curve tangent to $\bar{k}_a$.
Thus, the spatial part of $\bar{S}$ is associated with the time part as
\begin{align}
    \mathcal{S}(x^i_\text{o})-\mathcal{S}(x^i_\text{s}) = \omega(t_\text{o}-t_\text{s}).
\end{align}
$t_\text{o}-t_\text{s}$ is the time between the source and observer which can be obtained by solving the geodesic equation~\eqref{eq:geodesic}.

The first order of $\mathcal{B}$ of Eq.~\eqref{eq:spinopticsdispersionrelation} yields
\begin{align}
    \dot{x}_a \nabla^a \chi  &\approx \dot{x}_a \mathcal{B}^a.
    \label{eq:firstorderineom}
\end{align}
The right-hand side can be rewritten in terms of the language of gravito-electromagnetism and will be shown in the next section.

\subsubsection{\label{sec:GFRA}Alternative description with Gravito-electromagnetism}
Recall that the spin optics is equivalent to the gravitational Faraday rotation in the context of gravito-electromagnetism~\cite{Nouri-Zonoz:1999jls}.
The right-hand side Eq.~\eqref{eq:firstorderineom}, which is the result of spin optics, can be better understood and simply described with the gravito-electric field $\bm{E}$ and gravito-magnetic field $\bm{B}$ defined as 
\begin{align}
    \bm{E}:=&-\mathbf{D}\ln\sqrt{-g_{tt}}=-\frac{1}{2}\frac{\mathbf{D}g_{tt}}{g_{tt}},\\
    \bm{B}:=&\curl \bm{g}.
\end{align}
The operator $\curl$ is defined by
\begin{align}
    \curl g^i:=\epsilon^{ijk}\operatorname{D}_jg_k=\epsilon^{ijk}\partial_jg_k,
\end{align}
where we use $\epsilon^{ijk}{{^{(3)}}\Gamma^{l}}_{jk}=0$ in the second equality.
$\epsilon^{ijk}=\epsilon^{[ijk]}$ is the spatial anti-symmetric tensor with $\epsilon^{123}=1/\sqrt{\gamma}$ and  $\epsilon_{123}=\sqrt{\gamma}$, where $\gamma:=\det(\gamma_{ij})$.
In stationary spacetime, the Einstein equation and Bianchi identities can be rewritten in the quasi-Maxwell form \cite{Landau:1975pou}
\begin{align}
    \div\bm{B}=&0,\\
    \curl\bm{E}=&0,\\
    \div\bm{E}=&-\left[\frac{1}{2}\left(\sqrt{-g_{tt}}\ \bm{B}\right)^2 + \bm{E}^2\right]\nonumber\\
    & - 16\pi g_{tt} \left(T_{tt}-\frac{1}{2}g_{tt}T\right),\\
    \curl\left(\sqrt{-g_{tt}}\bm{B}\right)=&2\bm{E}\times\left(\sqrt{-g_{tt}}\ \bm{B}\right) + 16\pi \sqrt{-g_{tt}} \bm{\mathcal{J}},\label{eq:Ampere}\\
    {^{(3)}}R^{ij}=&\mathrm{D}^i E^j + \Bigl[\left(\sqrt{-g_{tt}}\ B^i\right)\left(\sqrt{-g_{tt}}\ B^j\right) \nonumber \\
    &-\left(\sqrt{-g_{tt}}\ B\right)^2\gamma^{ij}\Bigr]+E^iE^j\nonumber\\
    &-8\pi g_{tt} \left(T^{ij}-\frac{1}{2}g^{ij}T\right),
\end{align}
where ${^{(3)}}R^{ij}$ is the spatial Ricci tensor associated with $\gamma_{ij}$, $T_{ab}$ is the stress energy tensor of a matter, $ \mathcal{J}^i: = T^i{}_t$, and $\div \bm{E}:=\mathrm{D}_i E^i$.
The gravito-electric field is sourced by the gravitational self energy and the matter energy.
The gravito-magnetic field is sourced by these currents.

We rewrite the right-hand side of Eq.~\eqref{eq:firstorderineom} in terms of the gravito-magnetic component $\bm{B}$.~\footnote{Taking the spatial metric as $\hat{\gamma}_{ij}:=\gamma_{ij}/(-g_{tt})$, the final expression consistent with Eq. (102) in Ref.~\cite{Frolov:2011mh}.}
\begin{align}
    \dot{x}_a \mathcal{B}^a =& i m^{*b} \dot{x}^a \nabla_a m_b \nonumber \\
    =&-\frac{i\xi_{(t)}^c \dot{x}_c}{\sqrt{-g_{tt}}}m^{*b}({e_3}^a + u^a)\nabla_a m_{b}\nonumber \\
    =&-\frac{i\xi_{(t)}^c \dot{x}_c}{\sqrt{-g_{tt}}}m^{*b}{e_3}^a\nabla_a m_{b}+\frac{i\xi_{(t)}^c \dot{x}_c}{g_{tt}}m^{*b}\xi_{(t)}^a\nabla_a m_{b}\nonumber \\
    =&-\frac{i\xi_{(t)}^c \dot{x}_c}{-g_{tt}} m^{*b}\xi_{(t)}^a\nabla_a m_{b}\nonumber \\
    =&-\frac{i\xi_{(t)}^c \dot{x}_c}{-g_{tt}} m^{*b} m^a\nabla_a \xi_{(t)b} \nonumber \\
    =&\frac{i\xi_{(t)}^c \dot{x}_c}{-g_{tt}} i e_1^{[a}e_2^{b]}\nabla_{a} \xi_{(t)b}\nonumber \\
    =&-\frac{\xi_{(t)}^c \dot{x}_c}{-2g_{tt}} \eta^{abcd}u_a {e_3}_b \nabla_{c} \xi_{(t)d}\nonumber \\
    =&\frac{1}{2} u_a \dot{x}_b \eta^{abcd}  \nabla_{c} u_d\nonumber \\
    =&\sqrt{-g_{tt}}\frac{1}{2}\dot{\bm{x}} \cdot \bm{B}.
    \label{eq:dotxB}
\end{align}
We have used Eq.~\eqref{eq:spatialnormalvector} in the second equality, 
Eq.~\eqref{eq:timelikenormalvector} in the third equality, 
Eqs.~\eqref{eq:circularvectorcondition}~\eqref{eq:FWtransportcircular} in the fourth equality, Eq.~\eqref{eq:Lietransportcircular} in the fourth to fifth line, Eq.~\eqref{eq:circularandbasevector} and $e_1^b e_1^c \nabla_{c} \xi_{(t)b}=0$ due to the anti-symmetry $\nabla_{c} \xi_{(t)b}=-\nabla_{b} \xi_{(t)c}$ which is found from the Killing equation $\pounds_{\xi_{(t)}}g_{ab}=0$ in the sixth equality, $\eta^{abcd}u_a e_3{}_b =2e_1^{[c}e_2^{d]}$ and $\nabla_a \eta_{bcdf}=0$ in the seventh equality, Eqs.~\eqref{eq:timelikenormalvector},~\eqref{eq:spatialnormalvector}\footnote{$\eta_{abcd}u^a e_1^b e_2^c e_3^d=1$ because of $g=-1$ for the tetrad $\{u,e_1,e_2,e_3\}$.}, and $\eta^{abcd}u_a u_b=0$ in the eighth equality, and $ u^a\eta_a{}^{bcd}\nabla_c u_d = \sqrt{-g_{tt}}u^a\eta_a{}^{bcd}\partial_c (u_d/\sqrt{-g_{tt}}) = \sqrt{-g_{tt}} \gamma^b{}_i\epsilon^{ijk}\partial_j (u_k/\sqrt{-g_{tt}}) =\sqrt{-g_{tt}} \gamma^b{}_i \epsilon^{ijk}\partial_j g_k = \sqrt{-g_{tt}}\gamma^b{}_iB^i$.
Therefore, the gravitational Faraday rotation angle can be written in a similar form to the Faraday rotation in electromagnetism as
\begin{align}
    \dot{x}_a \nabla^a \chi  &\approx \frac{\sqrt{-g_{tt}}}{2}\dot{\bm{x}} \cdot \bm{B}.
    \label{eq:GFR}
\end{align}
$\dot{x}_a \nabla^a$ means the directional derivative in the direction of the velocity.

\subsection{\label{sec:diffractiontheory}Diffraction theory}
\begin{figure}[t]
    \includegraphics[width=0.7\columnwidth]{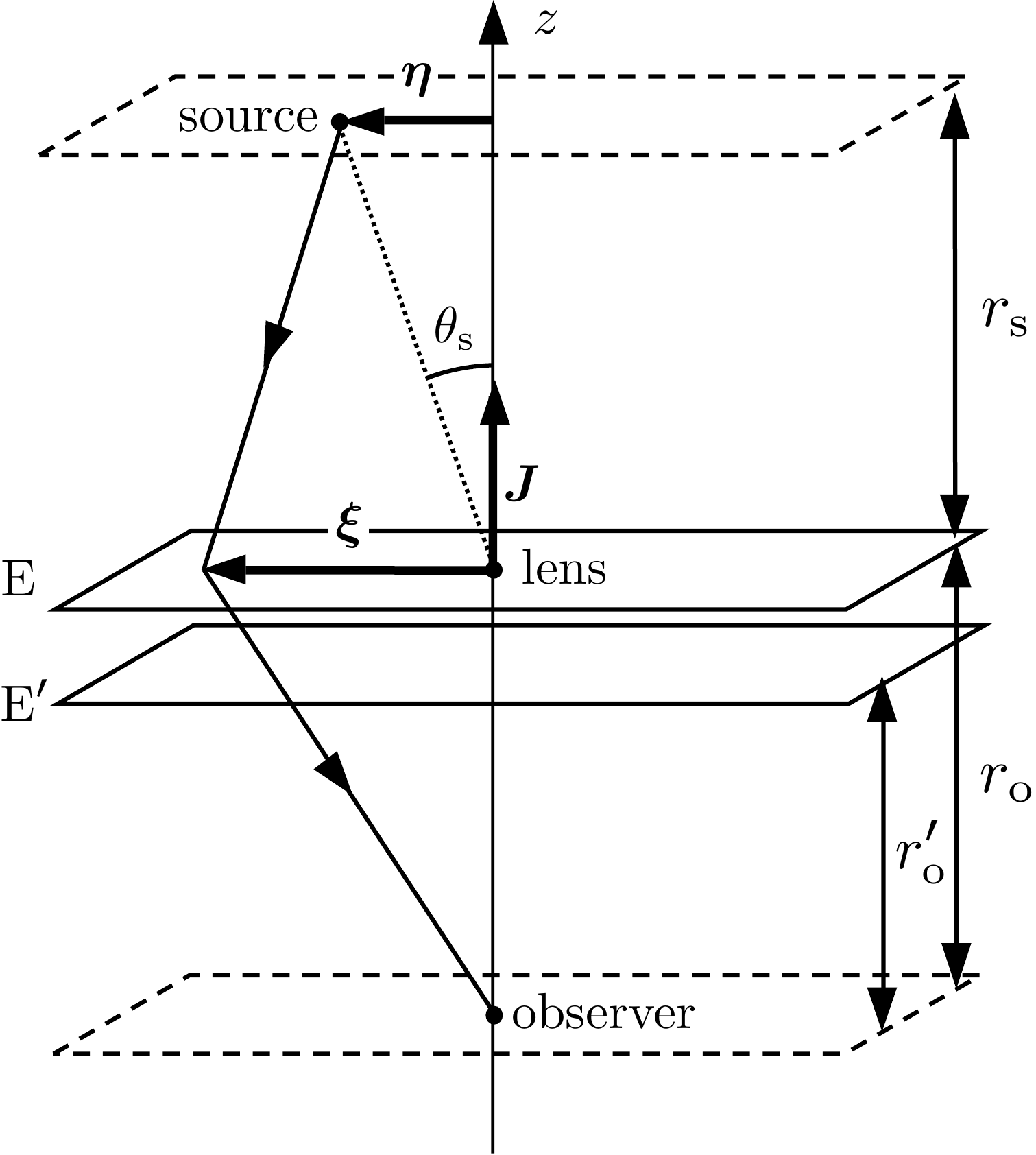}
    \caption{\label{fig:lens_geometry}Lens geometry on the $t=\text{const.}$ hypersurfaces. The true source position is at $\bm{\eta}$ with respect to the $z$-axis. The intersection point between the lens plane and the gravitational waves path at $\bm{\xi}$. $\text{E}$ is lens plane. $\text{E}'$ is intermediate plane. $r_\text{s}$ is the distance of the source from the lens plane $\text{E}$. $r_\text{o}$ and $r_\text{o}'$ are the distance of the observer from the lens plane $\text{E}$ and the intermediate plane $\text{E}'$, respectively. The lens plane $\text{E}$ is close to the intermediate plane $\text{E}'$ in the meaning of $r_\text{o}-r_\text{o}' \ll r_\text{o}$.
    }
\end{figure}
We have analytically obtained the phase incorporating the spin effect.
In this section, we plug the solution of the phase in the diffraction formula, and evaluate the arrival time and the magnification.

First, we introduce the diffraction formula, based on Ref.\cite{1992grle.book}.
Next, we evaluate the integral in the diffraction formula on the lens plane in the diffraction formula~\eqref{eq:diffractionformula} using the stationary phase approximation which can be applied to the gravitational wave with a wavelength shorter than the typical length scale of the lens.
In this paper, we configure the position of the source, lens, and observer shown in Fig.~\ref{fig:lens_geometry} so that the spin effect is at maximum.
We also assume that the background spacetime is asymptotically flat in the region far from the lens and the source is not too far from $z$-axis, i.e., $\eta \lesssim \xi$.

\subsubsection{\label{sec:diffractionformula}Diffraction formula}
The diffraction theory of gravitational lensing is well organized in \S 4.7 of Ref.~\cite{1992grle.book}.
In this book, the authors applied the diffraction theory to the intermediate plane $\text{E}'$ to the observer in which the background space-time is approximately flat.
The lens configuration and the spin effect which originated from the dragged component of the background spacetime do not change the derivation of the diffraction formula of gravitational lensing.
Therefore, the standard diffraction formula of gravitational lensing~\cite{Baraldo:1999ny,1992grle.book,TakahashiD} can be applied to the case in this paper, which is given by
\begin{align}
    \frac{\mathcal{E}_\text{obs}(\bm{\eta},\sigma)}{\mathcal{E}_\text{unlens}}=\frac{(r_\text{s}+r_\text{o})\omega }{2\pi i r_\text{s}r_\text{o}}\int_\text{E}\mathrm{d}^2\xi e^{i\mathcal{S}(\bm{\xi},\bm{\eta})}
    \label{eq:diffractionformula}
\end{align}
where $\mathcal{E}_\text{obs}$ and $\mathcal{E}_\text{unlens}$ are the amplitude at the observer with and without the lens, respectively.
The effect of spacetime distortion is imprinted only in the phase $\mathcal{S}$.

\subsubsection{\label{sec:stationalphaseapprox}stationary phase approximation}
We evaluate the integral of Eq.~\eqref{eq:diffractionformula} using stationary phase approximation.
In the stationary approximation, the third or higher-order Taylor expansion of the phase $\mathcal{S}$ at the stationary point $\bm{\Xi}$, which is the solution of $\partial_\xi S(\bm{\Xi}) = 0$, is assumed to be much smaller than the second-order Taylor expansion and hence neglect the higher-order terms.
This assumption is quantitatively translated into $\xi/\eta \ll M\omega$~\cite{Nakamura:1999uwi}.
In such a situation, the integrals around two stationary points $\bm{\Xi}$, which consist of the minimum point $\bm{\Xi}_\text{min}$ and the saddle point $\bm{\Xi}_\text{sad}$, are dominant.
Performing the multiple Gauss integral, the result which is the sum of the contribution from the two stationary points (see e.g. Ref.~\cite{1992grle.book}) is written as
\begin{align}
    \frac{\mathcal{E}_\text{obs}}{\mathcal{E}_\text{unlens}}=\sum_{j=\text{sud,min}}&\frac{(r_\text{s}+r_\text{o})\omega}{r_\text{s}r_\text{o}}\frac{\exp(i\mathcal{S}(\bm{\Xi}_j)-i\pi n_j/2)}{\sqrt{\left|\det(\bm{\nabla}\otimes\bm{\nabla}\mathcal{S}(\bm{\Xi_j}))\right|}},
\end{align}
where $n_\text{min}=0$, $n_\text{sad}=1$, $\otimes$ denotes tensor product and $\bm{\nabla}$ denotes the two-dimensional derivative on the lens plane.
$\bm{\nabla}\otimes\bm{\nabla}\mathcal{S}(\bm{\Xi_j})$ is the Hesse matrix at the stationary points.

\section{\label{sec:examplekerr}Example: Kerr black hole}
We apply our formalism to gravitational waves lensed by a Kerr black hole as an example.
Part of the calculation follows Ref.~\cite{Baraldo:1999ny}.
We consider the lens geometry drawn in Fig.~\ref{fig:lens_geometry}.
First, we introduce the phase.
Next, we put the phase into the diffraction formula~\eqref{eq:diffractionformula}.
Finally, we evaluate the integration using the stationary phase approximation and estimate the arrival time delay and magnification difference between left- and right-handed gravitational waves.

Here we introduce a bookkeeping parameter $\epsilon$ as
\begin{align}
    \frac{M}{\xi} = \mathcal{O}(\epsilon),
\end{align}
where $\xi$ is typically in the same order as the Einstein radius defined by
\begin{align}
    \xi_\text{E}:=\sqrt{\frac{4Mr_\text{s}r_\text{o}}{r_\text{s}+r_\text{o}}}.
    \label{eq:defofEinsteinradius}
\end{align}
We assume the angular momentum per mass $a$ ,i.e. a Kerr parameter, is in the same order as $M$, then
\begin{align}
    \frac{a}{\xi} = \mathcal{O}(\epsilon).
\end{align}
In addition, we assume $r_\text{s}$ is also in the same order as $r_\text{o}$.
Then Eq.~\eqref{eq:defofEinsteinradius} yield
\begin{align}
    \frac{\xi}{r_\text{s}} \sim \frac{\xi}{r_\text{o}} = \mathcal{O}(\epsilon).
\end{align}
To summarize, we assume the relation
\begin{align}
    \frac{M}{\xi} \sim \frac{a}{\xi} \sim \frac{\xi}{r_\text{s}} \sim \frac{\xi}{r_\text{o}} = \mathcal{O}(\epsilon).
    \label{eq:bookkeeping}
\end{align}
Because the $\chi \sim \mathcal{O}(\epsilon^3)$, we calculate it up to the third order of $\epsilon$.
Hereafter, ``$\simeq$'' denotes equality up to the third order of $\epsilon$.

\subsection{\label{sec:magneticcomponent}magnetic component}
The explicit expression of the gravito-magnetic components for Kerr metric in the Boyer-Lindquist coordinate is
\begin{align}
    \bm{B}=&-\frac{2aMr\Delta\sin(2\theta)}{\sqrt{\gamma}(\rho^2-2Mr)^2}\bm{\partial}_r\nonumber \\
    &-\frac{2aM(r^2-a^2\cos^2\theta)\sin^2\theta}{\sqrt{\gamma}(\rho^2-2Mr)^2}\bm{\partial}_\theta,
\end{align}
and its infinitesimal circulation is
\begin{align}
    \curl (\sqrt{-g_{tt}}\bm{B})= \frac{4 a M^2}{\rho^3 \left(\rho^2-2Mr\right)^{3/2}}\bm{\partial}_\phi,
    \label{eq:curlB}
\end{align}
where $\rho$ and $\Delta$ is defined in Eq.~\eqref{eq:defrho} and Eq~\eqref{eq:defDelta}, respectively.
Eq.~\eqref{eq:curlB} is satisfied Eq.~\eqref{eq:Ampere}.
Since the charge of the gravitational field is the energy, the gravito-magnetic field sourced by circular energy current is analogous to the gravito-magnetic field sourced by circular electric current.

\subsection{\label{sec:phase}phase}

We calculate the phase separating the helicity-independent and helicity-dependent parts based on the procedure described in \S~\ref{sec:dispersionrelation}.

First, we get the helicity-independent part of the phase. 
The covariant derivative Eq.~\eqref{eq:zerothorderofdisperaisionrelation} leads to the geometric equation
\begin{align}
    \bar{k}^b\nabla_b\bar{k}^a=0,
\end{align}
where $\bar{k}_a:=\nabla_a \bar{S}$.
Thus, $\bar{\mathcal{S}}$ is given by the time from the source to the observer multiplied by the frequency\footnote{Setting $k^a = \frac{\mathrm{d}x^a}{\mathrm{d}\nu}$, the null condition $\bar{k}^a \bar{k}_a = \frac{\mathrm{d}x^a}{\mathrm{d}\nu}\partial_a S = 0$ imply $\omega \mathrm{d}t = \mathrm{d}\mathcal{S}$.},
\begin{align}
    \bar{\mathcal{S}}=\omega \int_\text{sou}^\text{obs}\mathrm{d}t=\omega (t_\text{obs}-t_\text{sou}).
\end{align}
Using the null geodesic equations \eqref{eq:geodesicequationofr}-\eqref{eq:geodesicequationofphi},
\begin{align}
    &t_\text{o}-t_\text{s}\nonumber \\
    =&\left( \int_\text{sou}^\text{E} + \int_\text{E}^\text{obs}\right) \frac{r^2(r^2+a^2)+2aMr(a-L/E)}{\operatorname{sgn}(\bar{k}^r)\Delta \sqrt{R}}\mathrm{d}r \nonumber \\
    &+\int_\text{sou}^\text{obs}\frac{a^2\cos^2\theta}{\sqrt{\Theta}}\mathrm{d}\theta,
\end{align}
where $E$ and $L$ are defined in Eq.~\eqref{eq:defE} and Eq.~\eqref{eq:defL}, respectively.
The time is obtained in Ref.~\cite{Bray:1985ew} up to $\mathcal{O}(\epsilon^2)$.
Because the contribution of the rotation of the Kerr black hole in the time is $\mathcal{O}(\epsilon^2)$~\cite{Baraldo:1999ny}, the time is the same as in Schwarzschild black hole up to $\mathcal{O}(\epsilon^1)$.
The time up to $\mathcal{O}(\epsilon^1)$ is given by the summation of the Euclidean part and the Shapiro time delay.
Therefore, the eikonal is written as 
\begin{align}
    \bar{\mathcal{S}}(\bm{x},\bm{y})
    \simeq & \biggl[ \text{const.}\biggr] + \omega\biggl[ 2M|\bm{x}-\bm{y}|^2-4M\log(x) \biggr] \nonumber \\
    &+\biggl[\bar{\mathcal{S}}^{(2)}\biggr] + \biggl[\bar{\mathcal{S}}^{(3)}\biggr],
    \label{eq:totalphase}
\end{align}
where $\bm{x}$ and $\bm{y}$ are the dimensionless distance normalized the Einstein radius $\xi_\text{E}$ which are defined by
\begin{align}
    \bm{x}:=&\frac{\bm{\xi}}{\xi_\text{E}},&
    \bm{y}:=&\frac{\bm{\eta}}{\xi_\text{E}}\frac{r_\text{o}}{r_\text{s}+r_\text{o}},&
    \xi_\text{E}:=&\sqrt{\frac{4Mr_\text{o}r_\text{s}}{r_\text{s} + r_\text{o}}}.
\end{align}
The orders in the $\epsilon$ expansion of each square bracket are different and the square brackets are listed in lower order from the left.
$\bar{\mathcal{S}}^{(2)}$ and $\bar{\mathcal{S}}^{(3)}$ are not written down because they are not relevant to the main results in this paper, although they can be computed by performing the procedure described in Ref.~\cite{Bray:1985ew}.

Next, we calculate the helicity-independent part of the phase.
It is the first-order solution $\sigma \chi$ in $\mathcal{B}$ of the dispersion relation.~\eqref{eq:spinopticsdispersionrelation}.
$\chi$ is the gravitational Faraday rotation angle which is the solution of Eq.~\eqref{eq:GFR}.
The gravitational Faraday rotation angle for a Kerr black hole is calculated in App.~\ref{sec:gfrainKerr} (see also Ref.~\cite{Nouri-Zonoz:1999jls}), which is given by
\begin{align}
    \chi \simeq -\frac{\pi aM^2}{4\xi^3}.
\end{align}

Finally, we sum them to obtain the solution of the dispersion relation.~\eqref{eq:spinopticsdispersionrelation}, which is given by
\begin{align}
    \mathcal{S}(\bm{x},\bm{y})
    \simeq & \biggl[ \text{const.}\biggr] + \omega\biggl[ 2M|\bm{x}-\bm{y}|^2-4M\log(x) \biggr] \nonumber \\
    &+\biggl[\bar{\mathcal{S}}^{(2)}\biggr] + \biggl[\bar{\mathcal{S}}^{(3)}-\sigma\frac{\pi a M^2 }{4 x^3 \xi_\text{E}^3}\biggr].
    \label{eq:totalphase}
\end{align}
The last term is an important result led by spin optics.

\subsection{\label{sec:stationary}stationary point}
\begin{figure}[t]
    \includegraphics[width=0.6\columnwidth]{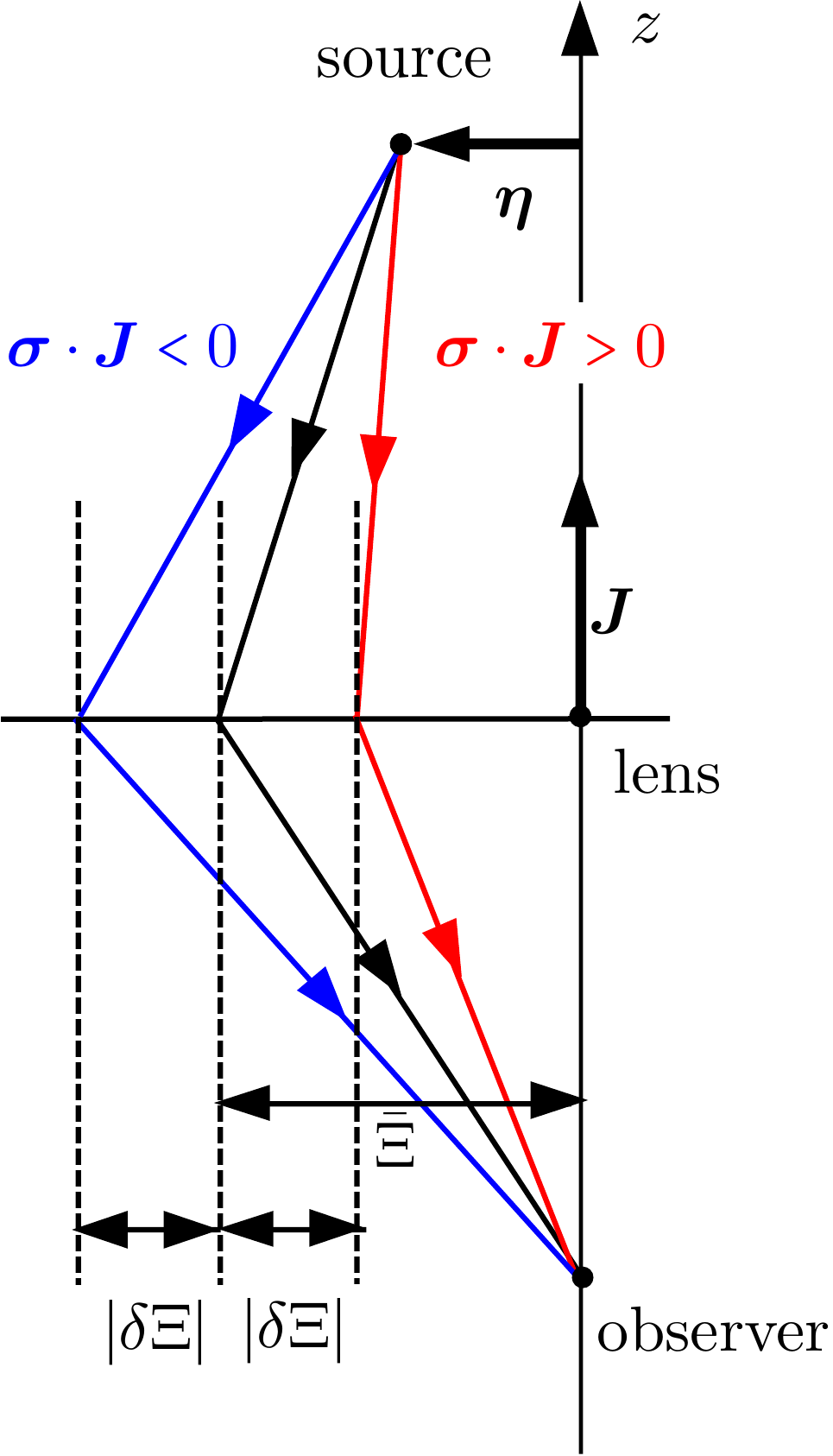}
    \caption{\label{fig:helicity_dependent_path}Schematic illustration of the trajectory of the spin-$0$ particle (black) and the gravitational waves with $\bm{\sigma}\cdot\bm{J}>0$ (red) and $\bm{\sigma}\cdot\bm{J}<0$ (blue).}
\end{figure}

The stationary points $\bm{x}=\bm{X}$ such that $\partial_{\bm{x}} S = 0$ are given by
\begin{widetext}
\begin{align}
    \bm{X}_\text{min}&\simeq\biggl[\frac{1}{2} \left(\sqrt{y^2+4}+y\right)\frac{\bm{y}}{y}\biggr]+\biggl[\bar{\bm{X}}_\text{min}^{(2)}\biggr]+\biggl[\bar{\bm{X}}_\text{min}^{(3)}-\sigma\frac{1}{\bar{X}_\text{min}^{(1)2} \left(\bar{X}_\text{min}^{(1)2}+1\right)}\frac{3 \pi  a M }{16 \omega \xi_\text{E}^3}\frac{\bm{y}}{y}\biggr],
    \label{eq:stationarypoint1}\\
    \bm{X}_\text{sad}&\simeq\biggl[\frac{1}{2} \left(\sqrt{y^2+4}-y\right)\frac{-\bm{y}}{y} \biggr]+\biggl[\bar{\bm{X}}_\text{sad}^{(2)}\biggr]+\biggl[\bar{\bm{X}}_\text{sad}^{(3)}-\sigma\frac{1}{\bar{X}_\text{sad}^{(1)2} \left(\bar{X}_\text{sad}^{(1)2}+1\right)}\frac{3 \pi  a M }{16 \omega \xi_\text{E}^3}\frac{-\bm{y}}{y}\biggr],
    \label{eq:stationarypoint2}
\end{align}
\end{widetext}
where
\begin{align}
    \bar{X}^{(1)}=\frac{1}{2} \left(\sqrt{y^2+4} \pm y \right).
\end{align}
The sign is $+$/$-$ for the minimum/saddle point.
$\bar{X}^{(2)}$ and $\bar{X}^{(3)}$ are not calculated in this paper.
The final term in Eqs.~\eqref{eq:stationarypoint1}~\eqref{eq:stationarypoint2} is helicity-dependent part due to the spin effect.
The stationary points (see also Fig.~\ref{fig:helicity_dependent_path}) are shifted from that of scalar waves by
\begin{align}
    \delta\Xi:=|\bm{X}-\bar{\bm{X}}|\xi_\text{E}=\frac{1}{\bar{X}^{(1)2} \left(\bar{X}^{(1)2}+1\right)}\frac{3 \pi  a M }{4 \omega \xi_\text{E}^2}.
\end{align}

\subsection{arrival time difference}
The gravitational Faraday rotation which is the last term in Eq.~\eqref{eq:totalphase} induces the arrival time difference between left- and right-handed gravitational waves.
Substituting the stationary points into the last term, we evaluate the arrival time delay between left- and right-handed gravitational waves
\begin{align}
    \delta t_\text{R-L} :=& \frac{\left|\mathcal{S}_\text{R}(\bm{X})-\mathcal{S}_\text{L}(\bm{X})\right|}{\omega}\nonumber \\
    \simeq & \frac{1}{\bar{X}^{(1)3}}\frac{2\pi}{\omega}\frac{a M^2 }{2\xi_\text{E}^3 }.
    \label{eq:timedelay}
\end{align}
Since $\delta t_\text{R-L} \propto 1/\omega$, the arrival time difference is large for long-wavelength waves.
The arrival difference between left-handed wave packet and right-handed wave packet calculated in Ref.~\cite{Harte:2022dpo,Oancea:2022szu} is $\delta t_\text{R-L} \propto 1/\omega^3$, which is different from the order in Eq.~\eqref{eq:timedelay}.
However, Ref.~\cite{Harte:2022dpo,Oancea:2022szu}, in which the author considers that a wave packet propagates near the Kerr black hole, is not an appropriate comparison since the situation is different from what we consider, i.e. the propagation of monochromatic waves. 
Therefore, it is not a problem if the frequency dependence is different from its result. 
As stated in introduction, gravitational waves in the regime of validity of linear perturbation can be decomposed into either a set of monochromatic waves or a set of wave packets, and the propagation of each element of the chosen basis can be studied separately. 
For a given astrophysical source of gravitational waves, the waveform at the position of the detector can be computed by taking an appropriate linear combination and should be independent of the choice of basis. Confirming this explicitly is, however, beyond the scope of the present paper and is left as a future work.

We estimate the arrival time difference for a hypothetical lens with the mass $M=10^{12} M_\odot$ and the Kerr parameter $a=M$ at the distance $r_\text{s}=r_\text{s}=D=\text{Gpc}$.
The factor $aM^2/\xi_\text{E}^3$ is $\mathcal{O}(10^{-9})$ and $\bar{X}^{(1)}$ is $\mathcal{O}(1)$.
Considering gravitational waves with the frequency $f=\omega/(2\pi)=\text{mHz}$ in the LISA band, the arrival time difference is estimated to be
\begin{align}
    \delta t_\text{R-L} \sim 10^{-6} \text{ sec }&\biggl(\frac{a}{10^{12}\text{ km}}\biggr)\biggl(\frac{M}{10^{12}\text{ km}}\biggr)^\frac{1}{2}\nonumber\\
    &\cdot\biggl(\frac{D}{\text{ Gpc}}\biggr)^{-\frac{3}{2}}\biggl(\frac{f}{\text{ mHz}}\biggr)^{-1}.
\end{align}

\subsection{\label{sec:magnification}magnification}
The normalized magnification defined by
\begin{align}
    \mu(\bm{y}) := \frac{(4M\omega)^2}{\det (\hat{\nabla} \otimes \hat{\nabla} \mathcal{S}(\bm{X},\bm{y}))},
\end{align}
where $\hat{\nabla}$ denotes the derivative with respect to the lens plane coordinate $\bm{x}$.
Those at the stationary points are explicitly
\begin{widetext}
\begin{align}
    |\mu_\text{min}|:=|\mu(\bm{X}_\text{min})|\simeq&\biggl[\frac{1}{2}+\frac{y^2+2}{2y\sqrt{y^2+4}}\biggr] + \biggl[\bar{\mu}_\text{min}^{(2)} \biggr]+\Biggl[\bar{\mu}_\text{min}^{(3)}+ \sigma\frac{ \bar{X}^{(1)}_\text{min}(3 \bar{X}^{(1)2}_\text{min}+1)}{(\bar{X}^{(1)2}_\text{min}-1) (\bar{X}^{(1)2}_\text{min}+1)^3 }\frac{3 \pi  a M }{16 \omega  \xi_\text{E}^3 }\Biggr],\\
    |\mu_\text{sad}|:=|\mu(\bm{X}_\text{sad})|\simeq&\biggl[\frac{1}{2}-\frac{y^2+2}{2y\sqrt{y^2+4}}\biggr] + \biggl[\bar{\mu}_\text{sad}^{(2)} \biggr]+\Biggl[\bar{\mu}_\text{sad}^{(3)}+ \sigma\frac{ \bar{X}^{(1)}_\text{sad}(3 \bar{X}^{(1)2}_\text{sad}+1)}{(\bar{X}^{(1)2}_\text{sad}-1) (\bar{X}^{(1)2}_\text{sad}+1)^3 }\frac{3 \pi  a M }{16 \omega  \xi_\text{E}^3 }\Biggr],\\
    \label{eq:magnificationforKerrBH}
\end{align}
\end{widetext}
where $\bar{\mu}^{(2)}$ and $\bar{\mu}^{(3)}$ are not calculated in this paper.
Only the final term depends on the helicity.
The magnification of gravitational waves with $\sigma=+2$ is more enhanced than that of scalar waves.
Intuitively, this is because gravitational waves with $\sigma=+2$ propagate over shorter distances than scalar waves (See Fig.~\ref{fig:helicity_dependent_path}).

The magnification difference between left- and right-handed gravitational waves is
\begin{align}
    \delta \mu_\text{R-L}:=&|\mu_\text{R}(\bm{X})-\mu_\text{L}(\bm{X})|\nonumber \\
    =&\frac{ \bar{X}^{(1)}(3 \bar{X}^{(1)2}+1)}{(\bar{X}^{(1)2}-1) (\bar{X}^{(1)2}+1)^3 }\frac{1}{M\omega}\frac{3 \pi  a M^2 }{4  \xi_\text{E}^3 }.
\end{align}
Since $\delta \mu_\text{R-L} \propto 1/\omega$, the magnification difference is also large for long-wavelength waves.
This means that when the source emits gravitational waves with pure plus-mode, i.e., $|\mathcal{E}_\text{R}|=|\mathcal{E}_\text{L}|$ at the source, the observer receives the elliptically polarized gravitational waves.
It is more elliptical for longer-wavelength gravitational waves.
However, since the factor $aM^2/\xi_\text{E}^3$ is very small $\lesssim \mathcal{O}(10^{-9})$ for wavelength shorter than the radius of curvature of a lens object, $\delta \mu_\text{R-L}$ is also small, where we assume a hypothetical lens with a mass $M=10^{12} M_\odot$ and the Kerr parameter $a=M$ at the distance $r_\text{s}=r_\text{s}=\text{Gpc}$.

The total amplitude is given by
\begin{align}
    \left|\frac{\mathcal{E}_\text{obs}}{\mathcal{E}_\text{unlens}}\right|^2 =& |\mu_\text{min}|+|\mu_\text{sad}|\nonumber \\
    &+2\sqrt{|\mu_\text{min}\mu_\text{sad}|}\sin(\mathcal{S}_\text{sad}-\mathcal{S}_\text{min}).
\end{align}
where $\mathcal{S}_\text{sad}:=\mathcal{S}(\bm{X}_\text{sad})$ and $\mathcal{S}_\text{min}:=\mathcal{S}(\bm{X}_\text{min})$.

\section{\label{sec:conclusion}conclusion}
We have developed the method to study the gravitational lensing of gravitational waves taking into account part of the spin effect of gravitational waves.
Combining the computational techniques developed in the literature for the spin effect of gravitational waves as well as the ones for gravitational lensing, we have developed for the first time a consistent computation for gravitational lensing incorporating the spin effect of the gravitational waves by using the diffraction formula.
We applied our formalism to the case of monochromatic gravitational waves lensed by a Kerr black hole conditioned by $M\omega \gg 1$, illuminated potential signatures of the spin-induced gravitational time delay and the elliptical polarizations for future gravitational-wave observations.
We have shown in the case of a Kerr black hole the time delay~\eqref{eq:timedelay} between left- and right-handed gravitational waves is more enhanced for longer-wavelength gravitational waves.
Furthermore, our results have also shown that the magnification~\eqref{eq:magnificationforKerrBH} depends on both the frequency and the helicity of gravitational waves.
For example, the magnification of gravitational waves with $\sigma = +2$ is larger than that of scalar waves in the lens geometry in Fig ~\ref{fig:lens_geometry}.
Notably, we have found that the difference of magnification between the left- and the right-handed gravitational waves is larger for longer wavelength gravitational waves.
Our predictions for the differences in phase and magnification could be observed in future gravitational-wave detectors, potentially giving new information on the rotation properties of a lens object and a dragged spacetime.

In this paper, we focus on gravitational waves only for $M\omega \gtrsim 1$.
To investigate the regime $M\omega \lesssim 1$, we need to calculate the diffraction integral without relying on the stationary phase approximation. 
Alternatively, one can employ the path integral method~\cite{Nakamura:1999uwi}, utilize the Teukolsky equation, or numerically solve Einstein equation~\cite{He:2022sjf}. These approaches are able to be extended to the range of $M\omega \gg 1$, making it worth comparing to our results of spin optics. In conclusion, studying the theoretical computation of wave scattering by a spinning object is still desirable in the future.

\begin{acknowledgments}
The work of S.M.~was supported in part by World Premier International Research Center Initiative (WPI), MEXT, Japan. S.M.~is grateful for the hospitality of Perimeter Institute where part of this work was carried out. Research at Perimeter Institute is supported in part by the Government of Canada through the Department of Innovation, Science and Economic Development and by the Province of Ontario through the Ministry of Colleges and Universities.
\end{acknowledgments}

\appendix
\section{\label{sec:basicequationsoskerr}Basic materials of Kerr spacetime}
We write down the basic equations in the Kerr spacetime based on \cite{Poisson:2009pwt}.
The Kerr metric in the Boyer-Lindquist coordinate is given by
\begin{align}
    \mathrm{d}s^2 =& -\left(1-\frac{2Mr}{\rho^2}\right)\mathrm{d}t^2 - \frac{4Mar\sin^2\theta}{\rho^2}\mathrm{d}\phi \mathrm{d}t \nonumber\\
    &+ \frac{\rho^2}{\Delta}\mathrm{d}r^2 + \rho^2\mathrm{d}\theta^2 +\frac{\Sigma}{\rho^2}\sin^2\theta\mathrm{d}\phi^2,
\end{align}
where 
\begin{align}
    \rho^2 &:= r^2 + a^2 \cos^2\theta, \label{eq:defrho}\\
    \Delta &:= r^2 - 2Mr + a^2, \label{eq:defDelta}\\
    \Sigma &:= (r^2 + a^2 )^2 - a^2 \Delta \sin^2 \theta.
\end{align}
The components of the inverse of the metric are
\begin{align}
    g^{ab}=\ 
    \begin{blockarray}{cccc}
    t & r & \theta  & \phi \\
     &  &  &  \\
    \begin{block}{(cccc)}
    \displaystyle -\frac{\Sigma}{\rho^2 \Delta} & 0 & 0  & \displaystyle -\frac{2Mar}{\rho^2 \Delta} \\
    0 & \displaystyle  \frac{\Delta}{\rho^2} & 0 & 0 \\
    0 & 0 & \displaystyle  \frac{1}{\rho^2}  & 0 \\
    \displaystyle  -\frac{2Mar}{\rho^2 \Sigma} & 0 & 0 & \displaystyle  \frac{\Delta - a^2 \sin^2\theta}{\rho^2 \Delta \sin^2\theta}\\
    \end{block}
    \end{blockarray}
    \ \ .
\end{align}

In the Kerr spacetime, there are two Killing vectors and one Killing tensor (See e.g. \cite{Isoyama:2013yor}) given by
\begin{align}
    \xi_{(t)}^a=&(\partial_t)^a=(1,0,0,0),\\
    \xi_{(\phi)}^a=&(\partial_\phi)^a=(0,0,0,1),\\
    \xi_{ab}=&2\Delta^2\mathfrak{l}_{(a}\mathfrak{n}_{b)}+r^2 g_{ab},
\end{align}
where $\mathfrak{l}$ and $\mathfrak{n}$ is defined by
\begin{align}
    \mathfrak{l}^a:=&(r^2 + a^2 ,\Delta, 0, a)/\Delta,\\
    \mathfrak{n}^a:=&(r^2 + a^2 ,-\Delta, 0, a)/\Delta.
\end{align}
The constants of motion such as the energy $E$, azimuthal angular momentum $L$, and Carter constant $\mathscr{Q}$, corresponding to each Killing vector and tensor exist as
\begin{align}
    E:=&-\xi_{(t)}^{a} p_a =-p_t,\label{eq:defE}\\
    L:=&\xi_{(\phi)}^{a} p_a = p_\phi,\label{eq:defL}\\
    \mathscr{Q}:=&\xi_{ab}p^a p^b-(L-aE)^2\label{eq:defQ}
\end{align}
In the case of monochromatic wave scattering that we consider in the paper, $E=\omega$.
For a null geodesic with a tangent vector $p^a=\mathrm{d}x^a/\mathrm{d}\nu$, writing Eqs. \eqref{eq:defE}-\eqref{eq:defQ} and $p^a p_a=0$ explicitly, we can obtain
\begin{align}
    \rho^2 \frac{\mathrm{d}t}{\mathrm{d}\nu}=&-a(aE\sin^2\theta-L)+(r^2+a^2)\frac{P}{\Delta},\label{eq:geodesicequationoft}\\
    \rho^2 \frac{\mathrm{d}r}{\mathrm{d}\nu}=&\pm E\sqrt{R},\label{eq:geodesicequationofr}\\
    \rho^2 \frac{\mathrm{d}\theta}{\mathrm{d}\nu}=&\pm E \sqrt{\Theta},\label{eq:geodesicequationoftheta}\\
    \rho^2 \frac{\mathrm{d}\phi}{\mathrm{d}\nu}=&-\left(aE-\frac{L}{\sin^2\theta}\right)+\frac{aP}{\Delta},\label{eq:geodesicequationofphi}
\end{align}
where
\begin{align}
    P(r):=&E(r^2+a^2)-aL,\\
    R(r)E^2:=&P^2-\Delta\bigl((L-aE)^2+\mathscr{Q}\bigr),\\
    \Theta(\theta)E^2 :=&\mathscr{Q}+\cos^2\theta\left(a^2E^2-\frac{L^2}{\sin^2\theta}\right).
\end{align}
The signs in Eqs.~\eqref{eq:geodesicequationofr}~\eqref{eq:geodesicequationoftheta} are positive for $\mathrm{d}r/\mathrm{d}\nu > 0, \mathrm{d}\theta/\mathrm{d}\nu > 0$, and negative for $\mathrm{d}r/\mathrm{d}\nu < 0, \mathrm{d}\theta/\mathrm{d}\nu < 0$, respectively.

\section{\label{sec:Commutation}Commutation relation}
We show that any Lie derivative along any Killing vector is commutative with any covariant derivative for any vector $V^a$.
$\nabla_a \Lie_\xi V_b$ and $\Lie_\xi \nabla_a V_b$ can be written as
\begin{align}
    \nabla_a \Lie_\xi V_b =& (\nabla_a \xi^c)(\nabla_c V_b) + \xi^c\nabla_a \nabla_c V_b \nonumber \\
    &+ (\nabla_a V_c)(\nabla_b \xi^c) + V_c\nabla_a \nabla_b \xi^c,\label{eq:commutation1}\\
    \Lie_\xi \nabla_a V_b =& \xi^c \nabla_c \nabla_a V_b + (\nabla_a \xi^c)(\nabla_c V_b) \nonumber \\
    &+ (\nabla_a V_c)(\nabla_b \xi^c).\label{eq:commutation2}
\end{align}
Here, we use the property of the Killing vector (see App. C.3 in Ref.~\cite{Wald:1984rg}),
\begin{align}
    \nabla_a \nabla_b \xi_c = -R_{bca}{}^d\xi_d.\label{eq:commutation7}
\end{align}
Using Eq.~\eqref{eq:commutation7}, one can show that the Riemann tensor that comes out when exchanging the covariant derivative of the second term on the right-hand side of Eq.~\eqref{eq:commutation1} cancels with that of the fourth term.
Then the right-hand side of \eqref{eq:commutation1} and Eq.~\eqref{eq:commutation2} are equivalent.
Therefore we obtain
\begin{align}
    [\Lie_\xi, \nabla_a] V_b = 0.
\end{align}

Performing the same procedure for any rank tensor, one can show any Lie derivative along any Killing vector is commutative with the covariant derivative.

\section{\label{sec:FermiWalkertransport}Fermi-Walker parallel transport}
We explain the Fermi-Walker parallel transport following~\cite{Hawking:1973uf,Misner:1973prb}.
The guiding principle of parallel transport is that the inner product of vectors transported along a curve is invariant.
Standard parallel transport of a vector $V^a$ along the integral curve of unit normal vector $U^a$ is the transport that satisfies
\begin{align}
    U^b \nabla_b V^a=0.
\end{align}
The standard parallel transported tangent vector along its geodesic curve is the tangent vector of the destination. However, the standard parallel transported tangent vector along the non-geodesic curve does not coincide with the tangent vector of the destination. The Fermi-Walker parallel transport is the transport such that (a) the inner product of transported vectors is invariant and (b) the transported tangent vector along the non-geodesic curve coincides with the tangent vector at the destination.
\par
The Fermi-Walker parallel transport is the transport with satisfying
\begin{align}
    \nabla^\text{FW}_U V^a := U^b \nabla_b V^a + \mathcal{F}^a{}_b V^b =0,
    \label{eq:FWtensor}
\end{align}
where
\begin{align}
    \mathcal{F}_{ab}=& \operatorname{sgn}(U^c U_c) \left(U_a A_b - A_a U_b \right),\nonumber \\
    A^a=&U^b\nabla_b U^a,
\end{align}
where $|U^a U_a|=1$.
Indeed, $\nabla^\text{FW}_U U^a=0$ and the derivative of the inner product between the tangent vector $U^a$ and Fermi-Walker transported vector $V^b$ along the curve is 
\begin{align}
    \frac{\mathrm{d}}{\mathrm{d}\tau} (U^a V_a)=&U^b\nabla_b (U^a V_a)=A^b V_a - U^a \mathcal{F}_a{}^b V_b \nonumber \\
    =&A^a V_a - \operatorname{sgn}(U^c U_c) U^a U_a  A^b V_b=0,
\end{align}
where we have used $U^a A_a=0$. 
The first term of Eq. \eqref{eq:FWtensor} is not necessary for the inner product with the tangent vector to be invariant but is necessary for the inner product between non-tangent vectors. The two transport for the geodesic curve, i.e. $A^a \propto U^a$, are equivalent because of $\mathcal{F}_{ab}=0$.
\begin{figure}[]
    \includegraphics[width=0.7\columnwidth]{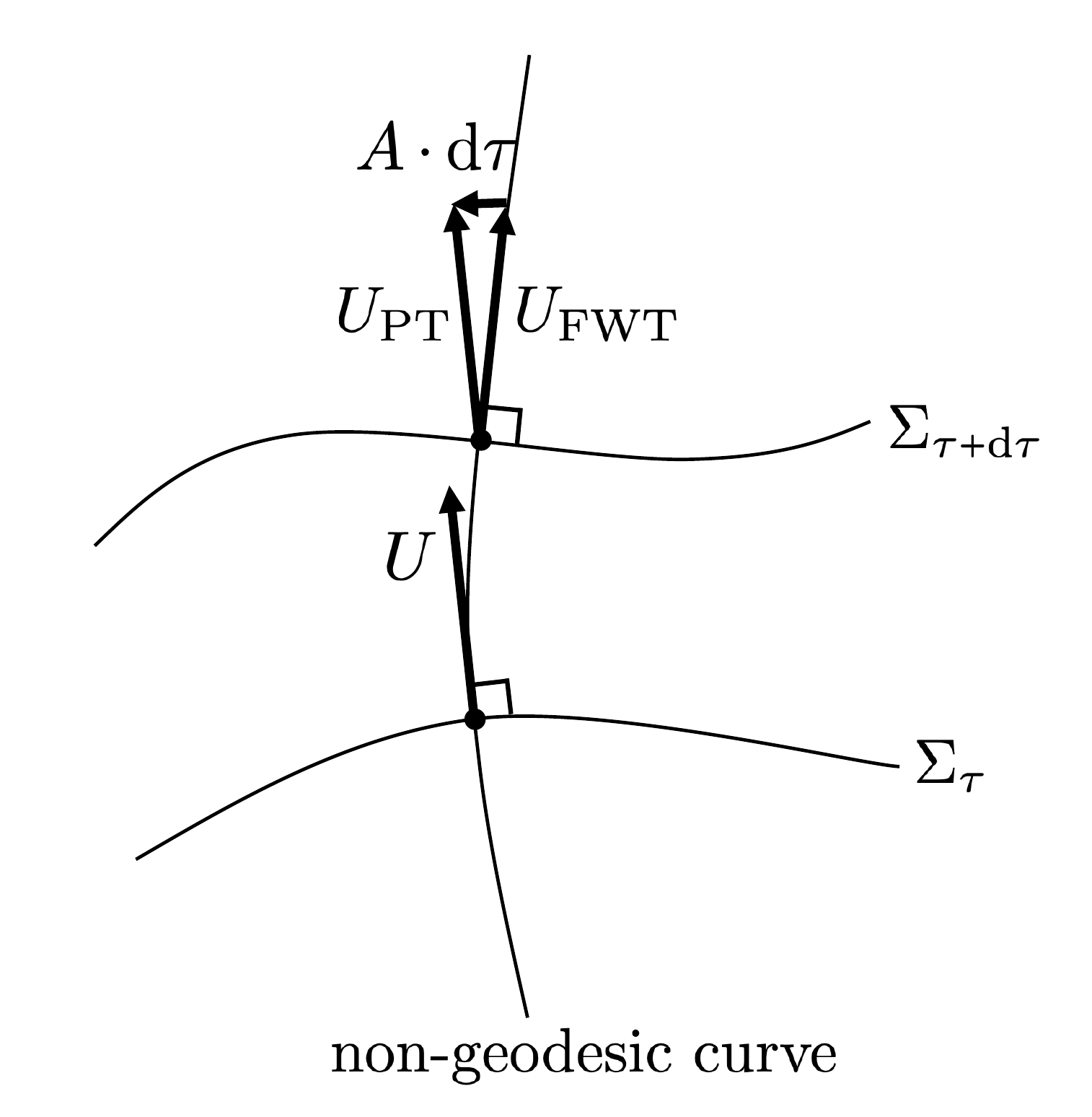}
    \caption{\label{fig:FWT}Standard parallel transported vector $U^a_\text{PT}$ and Fermi-Walker parallel transported vector $U^a_\text{FWT}$ of the tangent vector $U^a$ along the non-geodesic curve $U^b\nabla_b U^a = A^a$.}
\end{figure}

\section{\label{sec:gfrainKerr}Gravitational Faraday rotation angle in the Kerr spacetime}
Following Ref.~\cite{Nouri-Zonoz:1999jls}, we calculate the gravitational Faraday rotation angle $\chi$ in Kerr spacetime, which is defined by Eq.~\eqref{eq:GFR},
\begin{align}
    \dot{x}^a\nabla_a\chi = \frac{\sqrt{-g_{tt}}}{2}B_a \dot{x}^a = \frac{\sqrt{-g_{tt}}}{2}B_i \dot{x}^i.
    \label{eq:defofgfra}
\end{align}
Taking the parameter along the spatial path as $\lambda$ such that $\dot{x}^i\operatorname{D}_i:=\mathrm{d}/\mathrm{d}\lambda$, Eq.~\eqref{eq:defofgfra} yields
\begin{align}
    \mathrm{d}\chi=\frac{1}{2}\bm{B}\cdot\bm{e}_3\omega\mathrm{d}\lambda,
    \label{eq:infinitesimalchi}
\end{align}
where $e_3^i\simeq\sqrt{-g_{tt}}\dot{x}^i/\omega$.
$\dot{x}^a \dot{x}_a\simeq 0$~\eqref{eq:spinopticsdispersionrelation} leads to
\begin{align}
    \frac{\omega^2}{-g_{tt}}\simeq\left(\frac{\mathrm{d}l}{\mathrm{d}\lambda}\right)^2,
\end{align}
where $(\mathrm{d}l)^2\simeq\gamma_{ij}\mathrm{d}x^i\mathrm{d}x^j$ along the spatial path. 
This means $\bm{e}_3\omega\mathrm{d}\lambda \simeq \sqrt{-g_{tt}}\bm{e}_3\mathrm{d}l=:\sqrt{-g_{tt}}\mathrm{d}\bm{l}$, then Eq.~\eqref{eq:infinitesimalchi} is rewritten in the form
\begin{align}
    \chi \simeq &\frac{1}{2}\int\sqrt{-g_{tt}} \ \bm{B}\cdot \mathrm{d}\bm{l}.
    \label{eq:chi}
\end{align}
Since this expression is similar to the Faraday rotation angle in electromagnetic (see e.g.~\cite{Rybicki:2004hfl}), this is called ``gravitational Faraday rotation''.

\begin{figure}[t]
    \includegraphics[width=0.8\columnwidth]{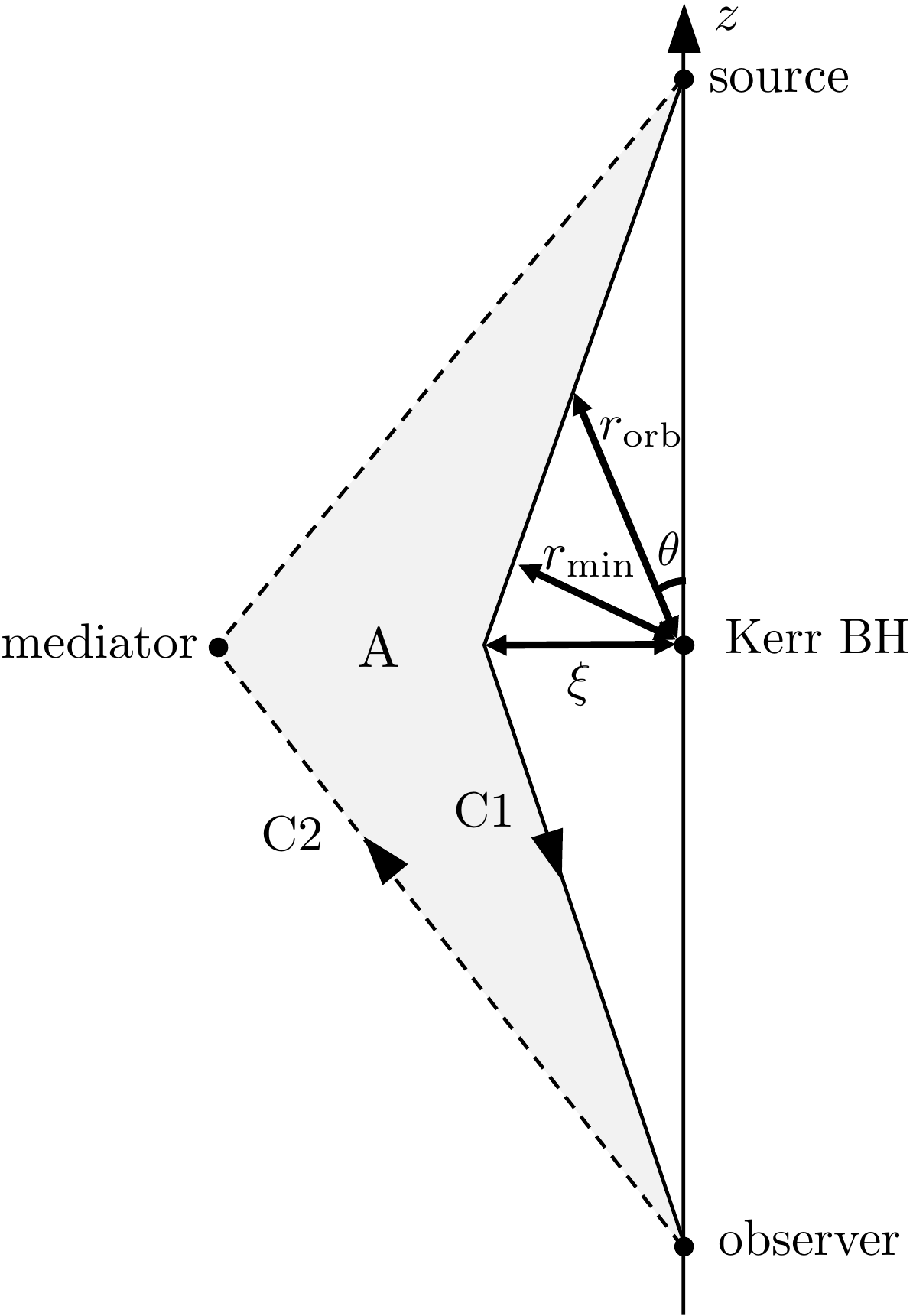}
    \caption{\label{fig:path}Path of integration.}
\end{figure}
First, we calculate $\chi$ for the case $\eta=0$.
Eq.~\eqref{eq:chi} into the surface integrate form for simplicity of calculation, we take close path $\text{C}=\text{C1}+\text{C2}$ drawn in Fig.~\ref{fig:path}~\footnote{The path $\text{C2}$ is different from one of Ref.~\cite{Nouri-Zonoz:1999jls}}.
The distance between the lens and the point on the path $\text{C2}$ is taken to be $\mathcal{O}(r_\text{s})$.
The path $\text{C1}$ routes near the lens, whereas path $\text{C2}$ is away from the lens.
Using the Stokes theorem, Eq.~\eqref{eq:chi} leads to
\begin{align}
    \frac{1}{2}\oint_\text{C}(\sqrt{-g_{tt}}\bm{B})\cdot \mathrm{d}\bm{l}=\frac{1}{2}\int_\text{S}\curl (\sqrt{-g_{tt}}\bm{B})\cdot\mathrm{d}\bm{S},
    \label{eq:stokes}
\end{align}
where $\text{S}$ denotes the area enclosed by the closed path $\text{C}$.
$\curl (\sqrt{-g_{tt}}\bm{B})$ for the Kerr spacetime is
\begin{align}
    \curl (\sqrt{-g_{tt}}\bm{B}) = \frac{4 a M^2}{\rho^3 \left(\rho^2-2Mr\right)^{3/2}}\bm{\partial}_\phi.
    \label{eq:curlBinKerr}
\end{align}
Because $r \curl (\sqrt{-g_{tt}}\bm{B})\to 0$ as $r \to \infty$, the integral along the path $\text{C2}$ is negligible~\footnote{The integral along the path $\text{C2}$ is $\mathcal{O}(aM/r_\text{s}^2)$}.
Therefore Eq.~\eqref{eq:stokes} becomes
\begin{align}
    \chi \simeq \frac{1}{2}\int_\text{C1}(\sqrt{-g_{tt}}\bm{B})\cdot \mathrm{d}\bm{l}=\frac{1}{2}\int_\text{S}\curl (\sqrt{-g_{tt}}\bm{B})\cdot\mathrm{d}\bm{A},
    \label{eq:chi2}
\end{align}
where $\mathrm{d}A^i=-\gamma^{ij}\epsilon_{jr\theta}\mathrm{d}r\mathrm{d}\theta\propto-(\partial_\phi)^i$ is the infinitesimal area of S.
The path $\text{C1}$ is bending due to gravity.
However, we can neglect the bending and the path $\text{C1}$ approximate with the straight line since the order of the integrand $\curl (\sqrt{-g_{tt}}\bm{B})\sim\mathcal{O}(\epsilon^3)$ is the highest order which we focus on.
Then Eqs.~\eqref{eq:curlBinKerr}~\eqref{eq:chi2} yield
\begin{align*}
    \chi &\simeq -2aM^2\int_{-1}^{1}\mathrm{d}\mu\int_{r_\text{orb}(\mu)}^{r_\text{C2}}\mathrm{d}r\frac{1}{r^4}\\
    &\simeq -\frac{2}{3}aM^2\int_{-1}^{1}\mathrm{d}\mu\frac{1}{r_\text{orb}^3}
\end{align*}
where $\mu:=\cos\theta$ and $r_\text{C2}$ is the distance to the path C2. Since $r_\text{C2}$ is the same order as $r_\text{s}$, we have neglected the term with $aM^2/r_\text{C2}^3=\mathcal{O}(\epsilon^6)$ in the second equality.
The relation between $r_\text{orb}$ and $\mu$ in the leading is 
\begin{align}
    r_\text{orb}\sim\frac{r_\text{min}}{\sqrt{1-\mu^2}}.
\end{align}
Thus, the gravitational Faraday rotation angle for Kerr spacetime is
\begin{align}
    \chi \simeq -\frac{\pi aM^2}{4r_\text{min}^3}.
    \label{eq:chi3}
\end{align}
This expression is consistent with Refs.~\cite{Nouri-Zonoz:1999jls}.
For the case $\xi\neq 0 $, $\chi$ is multiplied by the additional factor $\cos \theta_\mathrm{s}$~\cite{Nouri-Zonoz:1999jls}.
However, since $\cos \theta_\mathrm{s} = 1 + \mathcal{O}(\epsilon^2)$, the contribution of the additional factor is higher order and hence we neglect it.

\begin{figure}[t]
    \includegraphics[width=0.5\columnwidth]{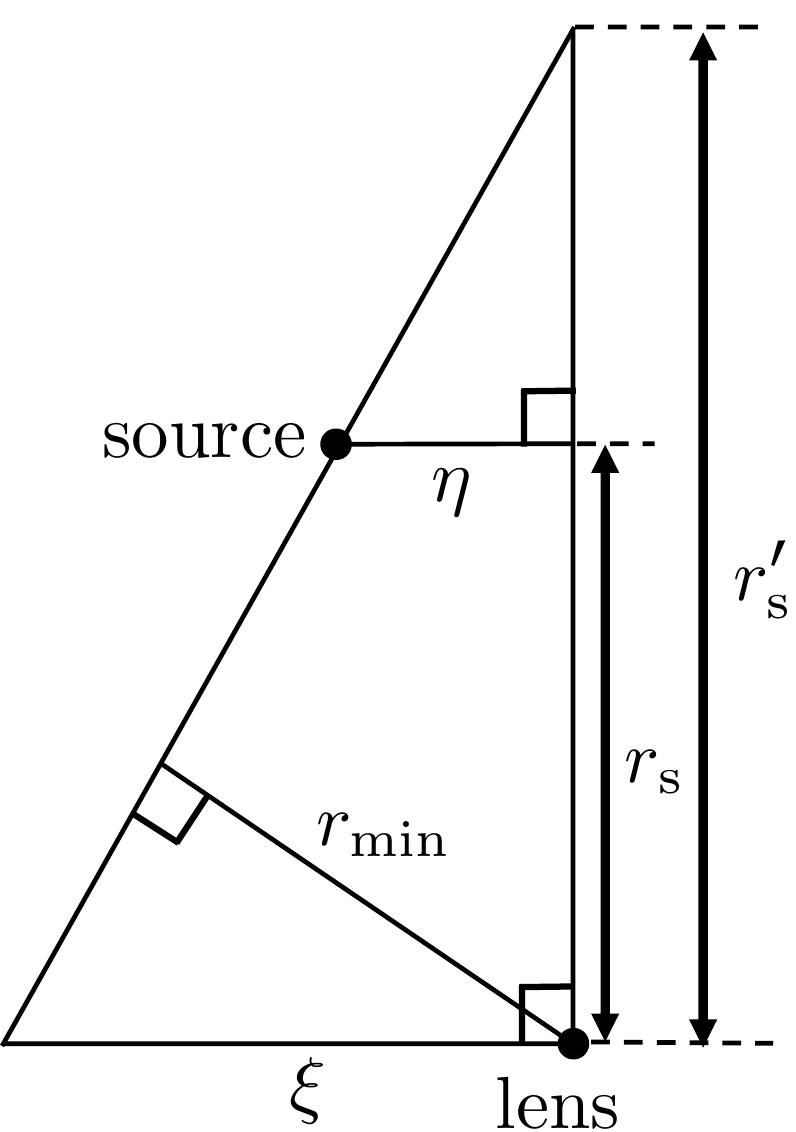}
    \caption{\label{fig:flat_geometry}Flat geometry. $r_\text{s}'$ is the distance from the lens to the intersection point between $z$ axis and the line extending the path.}
\end{figure}
Since the variable of integration in the diffraction formula is $\xi$, we rewrite Eq.~\eqref{eq:chi3} in terms of $\xi$.
Let us estimate the order of the difference between $r_\text{min}$ and $\xi$.
Here, we consider the path on the plane on which the source, lens, and observer are since the gravitational waves we focus on propagates on the plane in the leading order. 
The contribution of the distortion of spacetime in the length of $r_\text{min}$ and $\xi$ is higher order than the Euclidean part. 
Thus, we estimate it for the flat spacetime (Fig.~\ref{fig:flat_geometry}) in which the path is straight and we can use the Pythagorean theorem. 
The Pythagorean theorem yields $\xi / r_\text{min}=\sqrt{{r_\text{s}'}^2+\xi^2}/r_\text{s}'=1+\xi^2/(2{r_\text{s}'}^2)+\cdots.$
Using ${r_\text{s}'} > r_{\text{s}}$ and Eq.~\eqref{eq:bookkeeping}, we have $\xi / r_\text{min} - 1 < \mathcal{O}(\epsilon^2)$.
Since the order of $\chi$ is $\mathcal{O}(\epsilon^3)$, we can replace $r_\text{min}$ in Eq.~\eqref{eq:chi3} by $\xi$, as
\begin{align}
    \chi \simeq-\frac{\pi aM^2}{4\xi^3}.
    \label{eq:chi4}
\end{align}

\bibliographystyle{apsrev4-1}
\bibliography{references}

\begin{thebibliography}{66}%
\makeatletter
\providecommand \@ifxundefined [1]{%
 \@ifx{#1\undefined}
}%
\providecommand \@ifnum [1]{%
 \ifnum #1\expandafter \@firstoftwo
 \else \expandafter \@secondoftwo
 \fi
}%
\providecommand \@ifx [1]{%
 \ifx #1\expandafter \@firstoftwo
 \else \expandafter \@secondoftwo
 \fi
}%
\providecommand \natexlab [1]{#1}%
\providecommand \enquote  [1]{``#1''}%
\providecommand \bibnamefont  [1]{#1}%
\providecommand \bibfnamefont [1]{#1}%
\providecommand \citenamefont [1]{#1}%
\providecommand \href@noop [0]{\@secondoftwo}%
\providecommand \href [0]{\begingroup \@sanitize@url \@href}%
\providecommand \@href[1]{\@@startlink{#1}\@@href}%
\providecommand \@@href[1]{\endgroup#1\@@endlink}%
\providecommand \@sanitize@url [0]{\catcode `\\12\catcode `\$12\catcode
  `\&12\catcode `\#12\catcode `\^12\catcode `\_12\catcode `\%12\relax}%
\providecommand \@@startlink[1]{}%
\providecommand \@@endlink[0]{}%
\providecommand \url  [0]{\begingroup\@sanitize@url \@url }%
\providecommand \@url [1]{\endgroup\@href {#1}{\urlprefix }}%
\providecommand \urlprefix  [0]{URL }%
\providecommand \Eprint [0]{\href }%
\providecommand \doibase [0]{http://dx.doi.org/}%
\providecommand \selectlanguage [0]{\@gobble}%
\providecommand \bibinfo  [0]{\@secondoftwo}%
\providecommand \bibfield  [0]{\@secondoftwo}%
\providecommand \translation [1]{[#1]}%
\providecommand \BibitemOpen [0]{}%
\providecommand \bibitemStop [0]{}%
\providecommand \bibitemNoStop [0]{.\EOS\space}%
\providecommand \EOS [0]{\spacefactor3000\relax}%
\providecommand \BibitemShut  [1]{\csname bibitem#1\endcsname}%
\let\auto@bib@innerbib\@empty
\bibitem [{\citenamefont {Abbott}\ \emph
  {et~al.}(2021{\natexlab{a}})\citenamefont {Abbott} \emph
  {et~al.}}]{LIGOScientific:2021djp}%
  \BibitemOpen
  \bibfield  {author} {\bibinfo {author} {\bibfnamefont {R.}~\bibnamefont
  {Abbott}} \emph {et~al.} (\bibinfo {collaboration} {LIGO Scientific, VIRGO,
  KAGRA}),\ }\href@noop {} {\  (\bibinfo {year} {2021}{\natexlab{a}})},\
  \Eprint {http://arxiv.org/abs/2111.03606} {arXiv:2111.03606 [gr-qc]}
  \BibitemShut {NoStop}%
\bibitem [{\citenamefont {Amaro-Seoane}\ \emph {et~al.}(2017)\citenamefont
  {Amaro-Seoane} \emph {et~al.}}]{LISA:2017pwj}%
  \BibitemOpen
  \bibfield  {author} {\bibinfo {author} {\bibfnamefont {P.}~\bibnamefont
  {Amaro-Seoane}} \emph {et~al.} (\bibinfo {collaboration} {LISA}),\
  }\href@noop {} {\  (\bibinfo {year} {2017})},\ \Eprint
  {http://arxiv.org/abs/1702.00786} {arXiv:1702.00786 [astro-ph.IM]}
  \BibitemShut {NoStop}%
\bibitem [{\citenamefont {Kawamura}\ \emph {et~al.}(2006)\citenamefont
  {Kawamura} \emph {et~al.}}]{Kawamura:2006up}%
  \BibitemOpen
  \bibfield  {author} {\bibinfo {author} {\bibfnamefont {S.}~\bibnamefont
  {Kawamura}} \emph {et~al.},\ }\href {\doibase 10.1088/0264-9381/23/8/S17}
  {\bibfield  {journal} {\bibinfo  {journal} {Class. Quant. Grav.}\ }\textbf
  {\bibinfo {volume} {23}},\ \bibinfo {pages} {S125} (\bibinfo {year}
  {2006})}\BibitemShut {NoStop}%
\bibitem [{\citenamefont {Kawamura}\ \emph {et~al.}(2021)\citenamefont
  {Kawamura} \emph {et~al.}}]{Kawamura:2020pcg}%
  \BibitemOpen
  \bibfield  {author} {\bibinfo {author} {\bibfnamefont {S.}~\bibnamefont
  {Kawamura}} \emph {et~al.},\ }\href {\doibase 10.1093/ptep/ptab019}
  {\bibfield  {journal} {\bibinfo  {journal} {PTEP}\ }\textbf {\bibinfo
  {volume} {2021}},\ \bibinfo {pages} {05A105} (\bibinfo {year} {2021})},\
  \Eprint {http://arxiv.org/abs/2006.13545} {arXiv:2006.13545 [gr-qc]}
  \BibitemShut {NoStop}%
\bibitem [{\citenamefont {Nakamura}\ and\ \citenamefont
  {Deguchi}(1999)}]{Nakamura:1999uwi}%
  \BibitemOpen
  \bibfield  {author} {\bibinfo {author} {\bibfnamefont {T.~T.}\ \bibnamefont
  {Nakamura}}\ and\ \bibinfo {author} {\bibfnamefont {S.}~\bibnamefont
  {Deguchi}},\ }\href {\doibase 10.1143/ptps.133.137} {\bibfield  {journal}
  {\bibinfo  {journal} {Prog. Theor. Phys. Suppl.}\ }\textbf {\bibinfo {volume}
  {133}},\ \bibinfo {pages} {137} (\bibinfo {year} {1999})}\BibitemShut
  {NoStop}%
\bibitem [{\citenamefont {Takahashi}\ and\ \citenamefont
  {Nakamura}(2003)}]{Takahashi:2003ix}%
  \BibitemOpen
  \bibfield  {author} {\bibinfo {author} {\bibfnamefont {R.}~\bibnamefont
  {Takahashi}}\ and\ \bibinfo {author} {\bibfnamefont {T.}~\bibnamefont
  {Nakamura}},\ }\href {\doibase 10.1086/377430} {\bibfield  {journal}
  {\bibinfo  {journal} {Astrophys. J.}\ }\textbf {\bibinfo {volume} {595}},\
  \bibinfo {pages} {1039} (\bibinfo {year} {2003})},\ \Eprint
  {http://arxiv.org/abs/astro-ph/0305055} {arXiv:astro-ph/0305055} \BibitemShut
  {NoStop}%
\bibitem [{\citenamefont {Abbott}\ \emph
  {et~al.}(2017{\natexlab{a}})\citenamefont {Abbott} \emph
  {et~al.}}]{LIGOScientific:2017vwq}%
  \BibitemOpen
  \bibfield  {author} {\bibinfo {author} {\bibfnamefont {B.~P.}\ \bibnamefont
  {Abbott}} \emph {et~al.} (\bibinfo {collaboration} {LIGO Scientific,
  Virgo}),\ }\href {\doibase 10.1103/PhysRevLett.119.161101} {\bibfield
  {journal} {\bibinfo  {journal} {Phys. Rev. Lett.}\ }\textbf {\bibinfo
  {volume} {119}},\ \bibinfo {pages} {161101} (\bibinfo {year}
  {2017}{\natexlab{a}})},\ \Eprint {http://arxiv.org/abs/1710.05832}
  {arXiv:1710.05832 [gr-qc]} \BibitemShut {NoStop}%
\bibitem [{\citenamefont {Creminelli}\ and\ \citenamefont
  {Vernizzi}(2017)}]{Creminelli:2017sry}%
  \BibitemOpen
  \bibfield  {author} {\bibinfo {author} {\bibfnamefont {P.}~\bibnamefont
  {Creminelli}}\ and\ \bibinfo {author} {\bibfnamefont {F.}~\bibnamefont
  {Vernizzi}},\ }\href {\doibase 10.1103/PhysRevLett.119.251302} {\bibfield
  {journal} {\bibinfo  {journal} {Phys. Rev. Lett.}\ }\textbf {\bibinfo
  {volume} {119}},\ \bibinfo {pages} {251302} (\bibinfo {year} {2017})},\
  \Eprint {http://arxiv.org/abs/1710.05877} {arXiv:1710.05877 [astro-ph.CO]}
  \BibitemShut {NoStop}%
\bibitem [{\citenamefont {Ezquiaga}\ and\ \citenamefont
  {Zumalac\'arregui}(2017)}]{Ezquiaga:2017ekz}%
  \BibitemOpen
  \bibfield  {author} {\bibinfo {author} {\bibfnamefont {J.~M.}\ \bibnamefont
  {Ezquiaga}}\ and\ \bibinfo {author} {\bibfnamefont {M.}~\bibnamefont
  {Zumalac\'arregui}},\ }\href {\doibase 10.1103/PhysRevLett.119.251304}
  {\bibfield  {journal} {\bibinfo  {journal} {Phys. Rev. Lett.}\ }\textbf
  {\bibinfo {volume} {119}},\ \bibinfo {pages} {251304} (\bibinfo {year}
  {2017})},\ \Eprint {http://arxiv.org/abs/1710.05901} {arXiv:1710.05901
  [astro-ph.CO]} \BibitemShut {NoStop}%
\bibitem [{\citenamefont {Sakstein}\ and\ \citenamefont
  {Jain}(2017)}]{Sakstein:2017xjx}%
  \BibitemOpen
  \bibfield  {author} {\bibinfo {author} {\bibfnamefont {J.}~\bibnamefont
  {Sakstein}}\ and\ \bibinfo {author} {\bibfnamefont {B.}~\bibnamefont
  {Jain}},\ }\href {\doibase 10.1103/PhysRevLett.119.251303} {\bibfield
  {journal} {\bibinfo  {journal} {Phys. Rev. Lett.}\ }\textbf {\bibinfo
  {volume} {119}},\ \bibinfo {pages} {251303} (\bibinfo {year} {2017})},\
  \Eprint {http://arxiv.org/abs/1710.05893} {arXiv:1710.05893 [astro-ph.CO]}
  \BibitemShut {NoStop}%
\bibitem [{\citenamefont {Baker}\ \emph {et~al.}(2017)\citenamefont {Baker},
  \citenamefont {Bellini}, \citenamefont {Ferreira}, \citenamefont {Lagos},
  \citenamefont {Noller},\ and\ \citenamefont {Sawicki}}]{Baker:2017hug}%
  \BibitemOpen
  \bibfield  {author} {\bibinfo {author} {\bibfnamefont {T.}~\bibnamefont
  {Baker}}, \bibinfo {author} {\bibfnamefont {E.}~\bibnamefont {Bellini}},
  \bibinfo {author} {\bibfnamefont {P.~G.}\ \bibnamefont {Ferreira}}, \bibinfo
  {author} {\bibfnamefont {M.}~\bibnamefont {Lagos}}, \bibinfo {author}
  {\bibfnamefont {J.}~\bibnamefont {Noller}}, \ and\ \bibinfo {author}
  {\bibfnamefont {I.}~\bibnamefont {Sawicki}},\ }\href {\doibase
  10.1103/PhysRevLett.119.251301} {\bibfield  {journal} {\bibinfo  {journal}
  {Phys. Rev. Lett.}\ }\textbf {\bibinfo {volume} {119}},\ \bibinfo {pages}
  {251301} (\bibinfo {year} {2017})},\ \Eprint
  {http://arxiv.org/abs/1710.06394} {arXiv:1710.06394 [astro-ph.CO]}
  \BibitemShut {NoStop}%
\bibitem [{\citenamefont {Arai}\ and\ \citenamefont
  {Nishizawa}(2018)}]{Arai:2017hxj}%
  \BibitemOpen
  \bibfield  {author} {\bibinfo {author} {\bibfnamefont {S.}~\bibnamefont
  {Arai}}\ and\ \bibinfo {author} {\bibfnamefont {A.}~\bibnamefont
  {Nishizawa}},\ }\href {\doibase 10.1103/PhysRevD.97.104038} {\bibfield
  {journal} {\bibinfo  {journal} {Phys. Rev. D}\ }\textbf {\bibinfo {volume}
  {97}},\ \bibinfo {pages} {104038} (\bibinfo {year} {2018})},\ \Eprint
  {http://arxiv.org/abs/1711.03776} {arXiv:1711.03776 [gr-qc]} \BibitemShut
  {NoStop}%
\bibitem [{\citenamefont {Amendola}\ \emph {et~al.}(2018)\citenamefont
  {Amendola}, \citenamefont {Kunz}, \citenamefont {Saltas},\ and\ \citenamefont
  {Sawicki}}]{Amendola:2017orw}%
  \BibitemOpen
  \bibfield  {author} {\bibinfo {author} {\bibfnamefont {L.}~\bibnamefont
  {Amendola}}, \bibinfo {author} {\bibfnamefont {M.}~\bibnamefont {Kunz}},
  \bibinfo {author} {\bibfnamefont {I.~D.}\ \bibnamefont {Saltas}}, \ and\
  \bibinfo {author} {\bibfnamefont {I.}~\bibnamefont {Sawicki}},\ }\href
  {\doibase 10.1103/PhysRevLett.120.131101} {\bibfield  {journal} {\bibinfo
  {journal} {Phys. Rev. Lett.}\ }\textbf {\bibinfo {volume} {120}},\ \bibinfo
  {pages} {131101} (\bibinfo {year} {2018})},\ \Eprint
  {http://arxiv.org/abs/1711.04825} {arXiv:1711.04825 [astro-ph.CO]}
  \BibitemShut {NoStop}%
\bibitem [{\citenamefont {Abbott}\ \emph
  {et~al.}(2017{\natexlab{b}})\citenamefont {Abbott} \emph
  {et~al.}}]{LIGOScientific:2017ycc}%
  \BibitemOpen
  \bibfield  {author} {\bibinfo {author} {\bibfnamefont {B.~P.}\ \bibnamefont
  {Abbott}} \emph {et~al.} (\bibinfo {collaboration} {LIGO Scientific,
  Virgo}),\ }\href {\doibase 10.1103/PhysRevLett.119.141101} {\bibfield
  {journal} {\bibinfo  {journal} {Phys. Rev. Lett.}\ }\textbf {\bibinfo
  {volume} {119}},\ \bibinfo {pages} {141101} (\bibinfo {year}
  {2017}{\natexlab{b}})},\ \Eprint {http://arxiv.org/abs/1709.09660}
  {arXiv:1709.09660 [gr-qc]} \BibitemShut {NoStop}%
\bibitem [{\citenamefont {Abbott}\ \emph {et~al.}(2019)\citenamefont {Abbott}
  \emph {et~al.}}]{LIGOScientific:2018dkp}%
  \BibitemOpen
  \bibfield  {author} {\bibinfo {author} {\bibfnamefont {B.~P.}\ \bibnamefont
  {Abbott}} \emph {et~al.} (\bibinfo {collaboration} {LIGO Scientific,
  Virgo}),\ }\href {\doibase 10.1103/PhysRevLett.123.011102} {\bibfield
  {journal} {\bibinfo  {journal} {Phys. Rev. Lett.}\ }\textbf {\bibinfo
  {volume} {123}},\ \bibinfo {pages} {011102} (\bibinfo {year} {2019})},\
  \Eprint {http://arxiv.org/abs/1811.00364} {arXiv:1811.00364 [gr-qc]}
  \BibitemShut {NoStop}%
\bibitem [{\citenamefont {Takeda}\ \emph {et~al.}(2021)\citenamefont {Takeda},
  \citenamefont {Morisaki},\ and\ \citenamefont {Nishizawa}}]{Takeda:2020tjj}%
  \BibitemOpen
  \bibfield  {author} {\bibinfo {author} {\bibfnamefont {H.}~\bibnamefont
  {Takeda}}, \bibinfo {author} {\bibfnamefont {S.}~\bibnamefont {Morisaki}}, \
  and\ \bibinfo {author} {\bibfnamefont {A.}~\bibnamefont {Nishizawa}},\ }\href
  {\doibase 10.1103/PhysRevD.103.064037} {\bibfield  {journal} {\bibinfo
  {journal} {Phys. Rev. D}\ }\textbf {\bibinfo {volume} {103}},\ \bibinfo
  {pages} {064037} (\bibinfo {year} {2021})},\ \Eprint
  {http://arxiv.org/abs/2010.14538} {arXiv:2010.14538 [gr-qc]} \BibitemShut
  {NoStop}%
\bibitem [{\citenamefont {Abbott}\ \emph
  {et~al.}(2021{\natexlab{b}})\citenamefont {Abbott} \emph
  {et~al.}}]{LIGOScientific:2020tif}%
  \BibitemOpen
  \bibfield  {author} {\bibinfo {author} {\bibfnamefont {R.}~\bibnamefont
  {Abbott}} \emph {et~al.} (\bibinfo {collaboration} {LIGO Scientific,
  Virgo}),\ }\href {\doibase 10.1103/PhysRevD.103.122002} {\bibfield  {journal}
  {\bibinfo  {journal} {Phys. Rev. D}\ }\textbf {\bibinfo {volume} {103}},\
  \bibinfo {pages} {122002} (\bibinfo {year} {2021}{\natexlab{b}})},\ \Eprint
  {http://arxiv.org/abs/2010.14529} {arXiv:2010.14529 [gr-qc]} \BibitemShut
  {NoStop}%
\bibitem [{\citenamefont {Hagihara}\ \emph {et~al.}(2019)\citenamefont
  {Hagihara}, \citenamefont {Era}, \citenamefont {Iikawa}, \citenamefont
  {Nishizawa},\ and\ \citenamefont {Asada}}]{Hagihara:2019ihn}%
  \BibitemOpen
  \bibfield  {author} {\bibinfo {author} {\bibfnamefont {Y.}~\bibnamefont
  {Hagihara}}, \bibinfo {author} {\bibfnamefont {N.}~\bibnamefont {Era}},
  \bibinfo {author} {\bibfnamefont {D.}~\bibnamefont {Iikawa}}, \bibinfo
  {author} {\bibfnamefont {A.}~\bibnamefont {Nishizawa}}, \ and\ \bibinfo
  {author} {\bibfnamefont {H.}~\bibnamefont {Asada}},\ }\href {\doibase
  10.1103/PhysRevD.100.064010} {\bibfield  {journal} {\bibinfo  {journal}
  {Phys. Rev. D}\ }\textbf {\bibinfo {volume} {100}},\ \bibinfo {pages}
  {064010} (\bibinfo {year} {2019})},\ \Eprint
  {http://arxiv.org/abs/1904.02300} {arXiv:1904.02300 [gr-qc]} \BibitemShut
  {NoStop}%
\bibitem [{\citenamefont {Pang}\ \emph {et~al.}(2020)\citenamefont {Pang},
  \citenamefont {Lo}, \citenamefont {Wong}, \citenamefont {Li},\ and\
  \citenamefont {Van Den~Broeck}}]{Pang:2020pfz}%
  \BibitemOpen
  \bibfield  {author} {\bibinfo {author} {\bibfnamefont {P.~T.~H.}\
  \bibnamefont {Pang}}, \bibinfo {author} {\bibfnamefont {R.~K.~L.}\
  \bibnamefont {Lo}}, \bibinfo {author} {\bibfnamefont {I.~C.~F.}\ \bibnamefont
  {Wong}}, \bibinfo {author} {\bibfnamefont {T.~G.~F.}\ \bibnamefont {Li}}, \
  and\ \bibinfo {author} {\bibfnamefont {C.}~\bibnamefont {Van Den~Broeck}},\
  }\href {\doibase 10.1103/PhysRevD.101.104055} {\bibfield  {journal} {\bibinfo
   {journal} {Phys. Rev. D}\ }\textbf {\bibinfo {volume} {101}},\ \bibinfo
  {pages} {104055} (\bibinfo {year} {2020})},\ \Eprint
  {http://arxiv.org/abs/2003.07375} {arXiv:2003.07375 [gr-qc]} \BibitemShut
  {NoStop}%
\bibitem [{\citenamefont {Takeda}\ \emph {et~al.}(2022)\citenamefont {Takeda},
  \citenamefont {Morisaki},\ and\ \citenamefont {Nishizawa}}]{Takeda:2021hgo}%
  \BibitemOpen
  \bibfield  {author} {\bibinfo {author} {\bibfnamefont {H.}~\bibnamefont
  {Takeda}}, \bibinfo {author} {\bibfnamefont {S.}~\bibnamefont {Morisaki}}, \
  and\ \bibinfo {author} {\bibfnamefont {A.}~\bibnamefont {Nishizawa}},\ }\href
  {\doibase 10.1103/PhysRevD.105.084019} {\bibfield  {journal} {\bibinfo
  {journal} {Phys. Rev. D}\ }\textbf {\bibinfo {volume} {105}},\ \bibinfo
  {pages} {084019} (\bibinfo {year} {2022})},\ \Eprint
  {http://arxiv.org/abs/2105.00253} {arXiv:2105.00253 [gr-qc]} \BibitemShut
  {NoStop}%
\bibitem [{\citenamefont {Frolov}\ and\ \citenamefont
  {Shoom}(2011)}]{Frolov:2011mh}%
  \BibitemOpen
  \bibfield  {author} {\bibinfo {author} {\bibfnamefont {V.~P.}\ \bibnamefont
  {Frolov}}\ and\ \bibinfo {author} {\bibfnamefont {A.~A.}\ \bibnamefont
  {Shoom}},\ }\href {\doibase 10.1103/PhysRevD.84.044026} {\bibfield  {journal}
  {\bibinfo  {journal} {Phys. Rev. D}\ }\textbf {\bibinfo {volume} {84}},\
  \bibinfo {pages} {044026} (\bibinfo {year} {2011})},\ \Eprint
  {http://arxiv.org/abs/1105.5629} {arXiv:1105.5629 [gr-qc]} \BibitemShut
  {NoStop}%
\bibitem [{\citenamefont {Yoo}(2012)}]{Yoo:2012vv}%
  \BibitemOpen
  \bibfield  {author} {\bibinfo {author} {\bibfnamefont {C.-M.}\ \bibnamefont
  {Yoo}},\ }\href {\doibase 10.1103/PhysRevD.86.084005} {\bibfield  {journal}
  {\bibinfo  {journal} {Phys. Rev. D}\ }\textbf {\bibinfo {volume} {86}},\
  \bibinfo {pages} {084005} (\bibinfo {year} {2012})},\ \Eprint
  {http://arxiv.org/abs/1207.6833} {arXiv:1207.6833 [gr-qc]} \BibitemShut
  {NoStop}%
\bibitem [{\citenamefont {Dolan}(2017)}]{Dolan:2017zgu}%
  \BibitemOpen
  \bibfield  {author} {\bibinfo {author} {\bibfnamefont {S.~R.}\ \bibnamefont
  {Dolan}},\ }\href {\doibase 10.1142/S0218271818430101} {\bibfield  {journal}
  {\bibinfo  {journal} {Int. J. Mod. Phys. D}\ }\textbf {\bibinfo {volume}
  {27}},\ \bibinfo {pages} {1843010} (\bibinfo {year} {2017})},\ \Eprint
  {http://arxiv.org/abs/1806.08617} {arXiv:1806.08617 [gr-qc]} \BibitemShut
  {NoStop}%
\bibitem [{\citenamefont {Dolan}(2018)}]{Dolan:2018ydp}%
  \BibitemOpen
  \bibfield  {author} {\bibinfo {author} {\bibfnamefont {S.~R.}\ \bibnamefont
  {Dolan}},\ }\href@noop {} {\  (\bibinfo {year} {2018})},\ \Eprint
  {http://arxiv.org/abs/1801.02273} {arXiv:1801.02273 [gr-qc]} \BibitemShut
  {NoStop}%
\bibitem [{\citenamefont {Frolov}\ and\ \citenamefont
  {Shoom}(2012)}]{Frolov:2012zn}%
  \BibitemOpen
  \bibfield  {author} {\bibinfo {author} {\bibfnamefont {V.~P.}\ \bibnamefont
  {Frolov}}\ and\ \bibinfo {author} {\bibfnamefont {A.~A.}\ \bibnamefont
  {Shoom}},\ }\href {\doibase 10.1103/PhysRevD.86.024010} {\bibfield  {journal}
  {\bibinfo  {journal} {Phys. Rev. D}\ }\textbf {\bibinfo {volume} {86}},\
  \bibinfo {pages} {024010} (\bibinfo {year} {2012})},\ \Eprint
  {http://arxiv.org/abs/1205.4479} {arXiv:1205.4479 [gr-qc]} \BibitemShut
  {NoStop}%
\bibitem [{\citenamefont {Souriau}(1974)}]{AIHPA_1974__20_4_315_0}%
  \BibitemOpen
  \bibfield  {author} {\bibinfo {author} {\bibfnamefont {J.-M.}\ \bibnamefont
  {Souriau}},\ }\href {http://www.numdam.org/item/AIHPA_1974__20_4_315_0/}
  {\bibfield  {journal} {\bibinfo  {journal} {Annales de l'institut Henri
  Poincar\'e. Section A, Physique Th\'eorique}\ }\textbf {\bibinfo {volume}
  {20}},\ \bibinfo {pages} {315} (\bibinfo {year} {1974})}\BibitemShut
  {NoStop}%
\bibitem [{\citenamefont {Saturnini}(1976)}]{saturnini:tel-01344863}%
  \BibitemOpen
  \bibfield  {author} {\bibinfo {author} {\bibfnamefont {P.}~\bibnamefont
  {Saturnini}},\ }\emph {\bibinfo {title} {{Un mod{\`e}le de particule {\`a}
  spin de masse nulle dans le champ de gravitation}}},\ \href
  {https://hal.science/tel-01344863} {\bibinfo {type} {Theses}},\ \bibinfo
  {school} {{Universit{\'e} de Provence}} (\bibinfo {year} {1976})\BibitemShut
  {NoStop}%
\bibitem [{\citenamefont {Duval}\ \emph {et~al.}(2019)\citenamefont {Duval},
  \citenamefont {Marsot},\ and\ \citenamefont {Sch\"ucker}}]{Duval:2018hzh}%
  \BibitemOpen
  \bibfield  {author} {\bibinfo {author} {\bibfnamefont {C.}~\bibnamefont
  {Duval}}, \bibinfo {author} {\bibfnamefont {L.}~\bibnamefont {Marsot}}, \
  and\ \bibinfo {author} {\bibfnamefont {T.}~\bibnamefont {Sch\"ucker}},\
  }\href {\doibase 10.1103/PhysRevD.99.124037} {\bibfield  {journal} {\bibinfo
  {journal} {Phys. Rev. D}\ }\textbf {\bibinfo {volume} {99}},\ \bibinfo
  {pages} {124037} (\bibinfo {year} {2019})},\ \Eprint
  {http://arxiv.org/abs/1812.03014} {arXiv:1812.03014 [gr-qc]} \BibitemShut
  {NoStop}%
\bibitem [{\citenamefont {Duval}\ and\ \citenamefont
  {Schucker}(2017)}]{Duval:2016hxo}%
  \BibitemOpen
  \bibfield  {author} {\bibinfo {author} {\bibfnamefont {C.}~\bibnamefont
  {Duval}}\ and\ \bibinfo {author} {\bibfnamefont {T.}~\bibnamefont
  {Schucker}},\ }\href {\doibase 10.1103/PhysRevD.96.043517} {\bibfield
  {journal} {\bibinfo  {journal} {Phys. Rev. D}\ }\textbf {\bibinfo {volume}
  {96}},\ \bibinfo {pages} {043517} (\bibinfo {year} {2017})},\ \Eprint
  {http://arxiv.org/abs/1610.00555} {arXiv:1610.00555 [gr-qc]} \BibitemShut
  {NoStop}%
\bibitem [{\citenamefont {{Mathisson}}(2010)}]{2010GReGr..42.1011M}%
  \BibitemOpen
  \bibfield  {author} {\bibinfo {author} {\bibfnamefont {M.}~\bibnamefont
  {{Mathisson}}},\ }\href {\doibase 10.1007/s10714-010-0939-y} {\bibfield
  {journal} {\bibinfo  {journal} {General Relativity and Gravitation}\ }\textbf
  {\bibinfo {volume} {42}},\ \bibinfo {pages} {1011} (\bibinfo {year}
  {2010})}\BibitemShut {NoStop}%
\bibitem [{\citenamefont {{Papapetrou}}(1951)}]{1951RSPSA.209..248P}%
  \BibitemOpen
  \bibfield  {author} {\bibinfo {author} {\bibfnamefont {A.}~\bibnamefont
  {{Papapetrou}}},\ }\href {\doibase 10.1098/rspa.1951.0200} {\bibfield
  {journal} {\bibinfo  {journal} {Proceedings of the Royal Society of London
  Series A}\ }\textbf {\bibinfo {volume} {209}},\ \bibinfo {pages} {248}
  (\bibinfo {year} {1951})}\BibitemShut {NoStop}%
\bibitem [{\citenamefont {{Dixon}}(1964)}]{1964NCim...34..317D}%
  \BibitemOpen
  \bibfield  {author} {\bibinfo {author} {\bibfnamefont {W.~G.}\ \bibnamefont
  {{Dixon}}},\ }\href {\doibase 10.1007/BF02734579} {\bibfield  {journal}
  {\bibinfo  {journal} {Il Nuovo Cimento}\ }\textbf {\bibinfo {volume} {34}},\
  \bibinfo {pages} {317} (\bibinfo {year} {1964})}\BibitemShut {NoStop}%
\bibitem [{\citenamefont {Dixon}(2015)}]{Dixon:2015vxa}%
  \BibitemOpen
  \bibfield  {author} {\bibinfo {author} {\bibfnamefont {W.~G.}\ \bibnamefont
  {Dixon}},\ }\href {\doibase 10.1007/978-3-319-18335-0_1} {\bibfield
  {journal} {\bibinfo  {journal} {Fund. Theor. Phys.}\ }\textbf {\bibinfo
  {volume} {179}},\ \bibinfo {pages} {1} (\bibinfo {year} {2015})}\BibitemShut
  {NoStop}%
\bibitem [{\citenamefont {Yamamoto}(2018)}]{Yamamoto:2017gla}%
  \BibitemOpen
  \bibfield  {author} {\bibinfo {author} {\bibfnamefont {N.}~\bibnamefont
  {Yamamoto}},\ }\href {\doibase 10.1103/PhysRevD.98.061701} {\bibfield
  {journal} {\bibinfo  {journal} {Phys. Rev. D}\ }\textbf {\bibinfo {volume}
  {98}},\ \bibinfo {pages} {061701} (\bibinfo {year} {2018})},\ \Eprint
  {http://arxiv.org/abs/1708.03113} {arXiv:1708.03113 [hep-th]} \BibitemShut
  {NoStop}%
\bibitem [{\citenamefont {Oancea}\ \emph {et~al.}(2020)\citenamefont {Oancea},
  \citenamefont {Joudioux}, \citenamefont {Dodin}, \citenamefont {Ruiz},
  \citenamefont {Paganini},\ and\ \citenamefont {Andersson}}]{Oancea:2020khc}%
  \BibitemOpen
  \bibfield  {author} {\bibinfo {author} {\bibfnamefont {M.~A.}\ \bibnamefont
  {Oancea}}, \bibinfo {author} {\bibfnamefont {J.}~\bibnamefont {Joudioux}},
  \bibinfo {author} {\bibfnamefont {I.~Y.}\ \bibnamefont {Dodin}}, \bibinfo
  {author} {\bibfnamefont {D.~E.}\ \bibnamefont {Ruiz}}, \bibinfo {author}
  {\bibfnamefont {C.~F.}\ \bibnamefont {Paganini}}, \ and\ \bibinfo {author}
  {\bibfnamefont {L.}~\bibnamefont {Andersson}},\ }\href {\doibase
  10.1103/PhysRevD.102.024075} {\bibfield  {journal} {\bibinfo  {journal}
  {Phys. Rev. D}\ }\textbf {\bibinfo {volume} {102}},\ \bibinfo {pages}
  {024075} (\bibinfo {year} {2020})},\ \Eprint
  {http://arxiv.org/abs/2003.04553} {arXiv:2003.04553 [gr-qc]} \BibitemShut
  {NoStop}%
\bibitem [{\citenamefont {Oancea}\ \emph {et~al.}(2022)\citenamefont {Oancea},
  \citenamefont {Stiskalek},\ and\ \citenamefont
  {Zumalac\'arregui}}]{Oancea:2022szu}%
  \BibitemOpen
  \bibfield  {author} {\bibinfo {author} {\bibfnamefont {M.~A.}\ \bibnamefont
  {Oancea}}, \bibinfo {author} {\bibfnamefont {R.}~\bibnamefont {Stiskalek}}, \
  and\ \bibinfo {author} {\bibfnamefont {M.}~\bibnamefont {Zumalac\'arregui}},\
  }\href@noop {} {\  (\bibinfo {year} {2022})},\ \Eprint
  {http://arxiv.org/abs/2209.06459} {arXiv:2209.06459 [gr-qc]} \BibitemShut
  {NoStop}%
\bibitem [{\citenamefont {Andersson}\ and\ \citenamefont
  {Oancea}(2023)}]{Andersson:2023bvw}%
  \BibitemOpen
  \bibfield  {author} {\bibinfo {author} {\bibfnamefont {L.}~\bibnamefont
  {Andersson}}\ and\ \bibinfo {author} {\bibfnamefont {M.~A.}\ \bibnamefont
  {Oancea}},\ }\href@noop {} {\  (\bibinfo {year} {2023})},\ \Eprint
  {http://arxiv.org/abs/2302.13634} {arXiv:2302.13634 [gr-qc]} \BibitemShut
  {NoStop}%
\bibitem [{\citenamefont {Andersson}\ \emph {et~al.}(2021)\citenamefont
  {Andersson}, \citenamefont {Joudioux}, \citenamefont {Oancea},\ and\
  \citenamefont {Raj}}]{Andersson:2020gsj}%
  \BibitemOpen
  \bibfield  {author} {\bibinfo {author} {\bibfnamefont {L.}~\bibnamefont
  {Andersson}}, \bibinfo {author} {\bibfnamefont {J.}~\bibnamefont {Joudioux}},
  \bibinfo {author} {\bibfnamefont {M.~A.}\ \bibnamefont {Oancea}}, \ and\
  \bibinfo {author} {\bibfnamefont {A.}~\bibnamefont {Raj}},\ }\href {\doibase
  10.1103/PhysRevD.103.044053} {\bibfield  {journal} {\bibinfo  {journal}
  {Phys. Rev. D}\ }\textbf {\bibinfo {volume} {103}},\ \bibinfo {pages}
  {044053} (\bibinfo {year} {2021})},\ \Eprint
  {http://arxiv.org/abs/2012.08363} {arXiv:2012.08363 [gr-qc]} \BibitemShut
  {NoStop}%
\bibitem [{\citenamefont {Oancea}\ \emph {et~al.}(2019)\citenamefont {Oancea},
  \citenamefont {Paganini}, \citenamefont {Joudioux},\ and\ \citenamefont
  {Andersson}}]{Oancea:2019pgm}%
  \BibitemOpen
  \bibfield  {author} {\bibinfo {author} {\bibfnamefont {M.~A.}\ \bibnamefont
  {Oancea}}, \bibinfo {author} {\bibfnamefont {C.~F.}\ \bibnamefont
  {Paganini}}, \bibinfo {author} {\bibfnamefont {J.}~\bibnamefont {Joudioux}},
  \ and\ \bibinfo {author} {\bibfnamefont {L.}~\bibnamefont {Andersson}},\
  }\href@noop {} {\  (\bibinfo {year} {2019})},\ \Eprint
  {http://arxiv.org/abs/1904.09963} {arXiv:1904.09963 [gr-qc]} \BibitemShut
  {NoStop}%
\bibitem [{\citenamefont {Harte}\ and\ \citenamefont
  {Oancea}(2022)}]{Harte:2022dpo}%
  \BibitemOpen
  \bibfield  {author} {\bibinfo {author} {\bibfnamefont {A.~I.}\ \bibnamefont
  {Harte}}\ and\ \bibinfo {author} {\bibfnamefont {M.~A.}\ \bibnamefont
  {Oancea}},\ }\href {\doibase 10.1103/PhysRevD.105.104061} {\bibfield
  {journal} {\bibinfo  {journal} {Phys. Rev. D}\ }\textbf {\bibinfo {volume}
  {105}},\ \bibinfo {pages} {104061} (\bibinfo {year} {2022})},\ \Eprint
  {http://arxiv.org/abs/2203.01753} {arXiv:2203.01753 [gr-qc]} \BibitemShut
  {NoStop}%
\bibitem [{\citenamefont {Oancea}\ \emph {et~al.}(2023)\citenamefont {Oancea},
  \citenamefont {Stiskalek},\ and\ \citenamefont
  {Zumalac\'arregui}}]{Oancea:2023hgu}%
  \BibitemOpen
  \bibfield  {author} {\bibinfo {author} {\bibfnamefont {M.~A.}\ \bibnamefont
  {Oancea}}, \bibinfo {author} {\bibfnamefont {R.}~\bibnamefont {Stiskalek}}, \
  and\ \bibinfo {author} {\bibfnamefont {M.}~\bibnamefont {Zumalac\'arregui}},\
  }\href@noop {} {\  (\bibinfo {year} {2023})},\ \Eprint
  {http://arxiv.org/abs/2307.01903} {arXiv:2307.01903 [gr-qc]} \BibitemShut
  {NoStop}%
\bibitem [{\citenamefont {Dahal}(2023)}]{Dahal:2023ncl}%
  \BibitemOpen
  \bibfield  {author} {\bibinfo {author} {\bibfnamefont {P.~K.}\ \bibnamefont
  {Dahal}},\ }\href@noop {} {\  (\bibinfo {year} {2023})},\ \Eprint
  {http://arxiv.org/abs/2301.08250} {arXiv:2301.08250 [gr-qc]} \BibitemShut
  {NoStop}%
\bibitem [{\citenamefont {Dolan}(2008)}]{Dolan:2008kf}%
  \BibitemOpen
  \bibfield  {author} {\bibinfo {author} {\bibfnamefont {S.~R.}\ \bibnamefont
  {Dolan}},\ }\href {\doibase 10.1088/0264-9381/25/23/235002} {\bibfield
  {journal} {\bibinfo  {journal} {Class. Quant. Grav.}\ }\textbf {\bibinfo
  {volume} {25}},\ \bibinfo {pages} {235002} (\bibinfo {year} {2008})},\
  \Eprint {http://arxiv.org/abs/0801.3805} {arXiv:0801.3805 [gr-qc]}
  \BibitemShut {NoStop}%
\bibitem [{\citenamefont {Leite}\ \emph {et~al.}(2017)\citenamefont {Leite},
  \citenamefont {Dolan},\ and\ \citenamefont {Crispino}}]{Leite:2017zyb}%
  \BibitemOpen
  \bibfield  {author} {\bibinfo {author} {\bibfnamefont {L.~C.~S.}\
  \bibnamefont {Leite}}, \bibinfo {author} {\bibfnamefont {S.~R.}\ \bibnamefont
  {Dolan}}, \ and\ \bibinfo {author} {\bibfnamefont {L.~C.~B.}\ \bibnamefont
  {Crispino}},\ }\href {\doibase 10.1016/j.physletb.2017.09.048} {\bibfield
  {journal} {\bibinfo  {journal} {Phys. Lett. B}\ }\textbf {\bibinfo {volume}
  {774}},\ \bibinfo {pages} {130} (\bibinfo {year} {2017})},\ \Eprint
  {http://arxiv.org/abs/1707.01144} {arXiv:1707.01144 [gr-qc]} \BibitemShut
  {NoStop}%
\bibitem [{\citenamefont {Leite}\ \emph {et~al.}(2018)\citenamefont {Leite},
  \citenamefont {Dolan},\ and\ \citenamefont {Crispino}}]{Leite:2018mon}%
  \BibitemOpen
  \bibfield  {author} {\bibinfo {author} {\bibfnamefont {L.~C.~S.}\
  \bibnamefont {Leite}}, \bibinfo {author} {\bibfnamefont {S.}~\bibnamefont
  {Dolan}}, \ and\ \bibinfo {author} {\bibfnamefont {L.}~\bibnamefont
  {Crispino}, \bibfnamefont {C.~B.}},\ }\href {\doibase
  10.1103/PhysRevD.98.024046} {\bibfield  {journal} {\bibinfo  {journal} {Phys.
  Rev. D}\ }\textbf {\bibinfo {volume} {98}},\ \bibinfo {pages} {024046}
  (\bibinfo {year} {2018})},\ \Eprint {http://arxiv.org/abs/1805.07840}
  {arXiv:1805.07840 [gr-qc]} \BibitemShut {NoStop}%
\bibitem [{\citenamefont {Tambalo}\ \emph {et~al.}(2022)\citenamefont
  {Tambalo}, \citenamefont {Zumalac\'arregui}, \citenamefont {Dai},\ and\
  \citenamefont {Cheung}}]{Tambalo:2022wlm}%
  \BibitemOpen
  \bibfield  {author} {\bibinfo {author} {\bibfnamefont {G.}~\bibnamefont
  {Tambalo}}, \bibinfo {author} {\bibfnamefont {M.}~\bibnamefont
  {Zumalac\'arregui}}, \bibinfo {author} {\bibfnamefont {L.}~\bibnamefont
  {Dai}}, \ and\ \bibinfo {author} {\bibfnamefont {M.~H.-Y.}\ \bibnamefont
  {Cheung}},\ }\href@noop {} {\  (\bibinfo {year} {2022})},\ \Eprint
  {http://arxiv.org/abs/2212.11960} {arXiv:2212.11960 [astro-ph.CO]}
  \BibitemShut {NoStop}%
\bibitem [{\citenamefont {Leung}\ \emph {et~al.}(2023)\citenamefont {Leung},
  \citenamefont {Jow}, \citenamefont {Saha}, \citenamefont {Dai}, \citenamefont
  {Oguri},\ and\ \citenamefont {Koopmans}}]{Leung:2023lmq}%
  \BibitemOpen
  \bibfield  {author} {\bibinfo {author} {\bibfnamefont {C.}~\bibnamefont
  {Leung}}, \bibinfo {author} {\bibfnamefont {D.}~\bibnamefont {Jow}}, \bibinfo
  {author} {\bibfnamefont {P.}~\bibnamefont {Saha}}, \bibinfo {author}
  {\bibfnamefont {L.}~\bibnamefont {Dai}}, \bibinfo {author} {\bibfnamefont
  {M.}~\bibnamefont {Oguri}}, \ and\ \bibinfo {author} {\bibfnamefont
  {L.~V.~E.}\ \bibnamefont {Koopmans}},\ }\href@noop {} {\  (\bibinfo {year}
  {2023})},\ \Eprint {http://arxiv.org/abs/2304.01202} {arXiv:2304.01202
  [astro-ph.HE]} \BibitemShut {NoStop}%
\bibitem [{\citenamefont {Baraldo}\ \emph {et~al.}(1999)\citenamefont
  {Baraldo}, \citenamefont {Hosoya},\ and\ \citenamefont
  {Nakamura}}]{Baraldo:1999ny}%
  \BibitemOpen
  \bibfield  {author} {\bibinfo {author} {\bibfnamefont {C.}~\bibnamefont
  {Baraldo}}, \bibinfo {author} {\bibfnamefont {A.}~\bibnamefont {Hosoya}}, \
  and\ \bibinfo {author} {\bibfnamefont {T.~T.}\ \bibnamefont {Nakamura}},\
  }\href {\doibase 10.1103/PhysRevD.59.083001} {\bibfield  {journal} {\bibinfo
  {journal} {Phys. Rev. D}\ }\textbf {\bibinfo {volume} {59}},\ \bibinfo
  {pages} {083001} (\bibinfo {year} {1999})}\BibitemShut {NoStop}%
\bibitem [{\citenamefont {Li}\ \emph {et~al.}(2022)\citenamefont {Li},
  \citenamefont {Qiao}, \citenamefont {Zhao},\ and\ \citenamefont
  {Er}}]{Li:2022izh}%
  \BibitemOpen
  \bibfield  {author} {\bibinfo {author} {\bibfnamefont {Z.}~\bibnamefont
  {Li}}, \bibinfo {author} {\bibfnamefont {J.}~\bibnamefont {Qiao}}, \bibinfo
  {author} {\bibfnamefont {W.}~\bibnamefont {Zhao}}, \ and\ \bibinfo {author}
  {\bibfnamefont {X.}~\bibnamefont {Er}},\ }\href {\doibase
  10.1088/1475-7516/2022/10/095} {\bibfield  {journal} {\bibinfo  {journal}
  {JCAP}\ }\textbf {\bibinfo {volume} {10}},\ \bibinfo {pages} {095} (\bibinfo
  {year} {2022})},\ \Eprint {http://arxiv.org/abs/2204.10512} {arXiv:2204.10512
  [gr-qc]} \BibitemShut {NoStop}%
\bibitem [{\citenamefont {Ishihara}\ \emph {et~al.}(1988)\citenamefont
  {Ishihara}, \citenamefont {Takahashi},\ and\ \citenamefont
  {Tomimatsu}}]{Ishihara:1987dv}%
  \BibitemOpen
  \bibfield  {author} {\bibinfo {author} {\bibfnamefont {H.}~\bibnamefont
  {Ishihara}}, \bibinfo {author} {\bibfnamefont {M.}~\bibnamefont {Takahashi}},
  \ and\ \bibinfo {author} {\bibfnamefont {A.}~\bibnamefont {Tomimatsu}},\
  }\href {\doibase 10.1103/PhysRevD.38.472} {\bibfield  {journal} {\bibinfo
  {journal} {Phys. Rev. D}\ }\textbf {\bibinfo {volume} {38}},\ \bibinfo
  {pages} {472} (\bibinfo {year} {1988})}\BibitemShut {NoStop}%
\bibitem [{\citenamefont {Nouri-Zonoz}(1999)}]{Nouri-Zonoz:1999jls}%
  \BibitemOpen
  \bibfield  {author} {\bibinfo {author} {\bibfnamefont {M.}~\bibnamefont
  {Nouri-Zonoz}},\ }\href {\doibase 10.1103/PhysRevD.60.024013} {\bibfield
  {journal} {\bibinfo  {journal} {Phys. Rev. D}\ }\textbf {\bibinfo {volume}
  {60}},\ \bibinfo {pages} {024013} (\bibinfo {year} {1999})},\ \Eprint
  {http://arxiv.org/abs/gr-qc/9901011} {arXiv:gr-qc/9901011} \BibitemShut
  {NoStop}%
\bibitem [{\citenamefont {Chakraborty}(2022)}]{Chakraborty:2021bsb}%
  \BibitemOpen
  \bibfield  {author} {\bibinfo {author} {\bibfnamefont {C.}~\bibnamefont
  {Chakraborty}},\ }\href {\doibase 10.1103/PhysRevD.105.064072} {\bibfield
  {journal} {\bibinfo  {journal} {Phys. Rev. D}\ }\textbf {\bibinfo {volume}
  {105}},\ \bibinfo {pages} {064072} (\bibinfo {year} {2022})},\ \Eprint
  {http://arxiv.org/abs/2106.03520} {arXiv:2106.03520 [gr-qc]} \BibitemShut
  {NoStop}%
\bibitem [{\citenamefont {Tamburini}\ \emph {et~al.}(2021)\citenamefont
  {Tamburini}, \citenamefont {Feleppa}, \citenamefont {Licata},\ and\
  \citenamefont {Thid\'e}}]{Tamburini:2021jok}%
  \BibitemOpen
  \bibfield  {author} {\bibinfo {author} {\bibfnamefont {F.}~\bibnamefont
  {Tamburini}}, \bibinfo {author} {\bibfnamefont {F.}~\bibnamefont {Feleppa}},
  \bibinfo {author} {\bibfnamefont {I.}~\bibnamefont {Licata}}, \ and\ \bibinfo
  {author} {\bibfnamefont {B.}~\bibnamefont {Thid\'e}},\ }\href {\doibase
  10.1103/PhysRevA.104.013718} {\bibfield  {journal} {\bibinfo  {journal}
  {Phys. Rev. A}\ }\textbf {\bibinfo {volume} {104}},\ \bibinfo {pages}
  {013718} (\bibinfo {year} {2021})},\ \Eprint
  {http://arxiv.org/abs/2104.06998} {arXiv:2104.06998 [gr-qc]} \BibitemShut
  {NoStop}%
\bibitem [{\citenamefont {Frolov}(2020)}]{Frolov:2020uhn}%
  \BibitemOpen
  \bibfield  {author} {\bibinfo {author} {\bibfnamefont {V.~P.}\ \bibnamefont
  {Frolov}},\ }\href {\doibase 10.1103/PhysRevD.102.084013} {\bibfield
  {journal} {\bibinfo  {journal} {Phys. Rev. D}\ }\textbf {\bibinfo {volume}
  {102}},\ \bibinfo {pages} {084013} (\bibinfo {year} {2020})},\ \Eprint
  {http://arxiv.org/abs/2007.03743} {arXiv:2007.03743 [gr-qc]} \BibitemShut
  {NoStop}%
\bibitem [{\citenamefont {Shoom}(2021)}]{Shoom:2020zhr}%
  \BibitemOpen
  \bibfield  {author} {\bibinfo {author} {\bibfnamefont {A.~A.}\ \bibnamefont
  {Shoom}},\ }\href {\doibase 10.1103/PhysRevD.104.084007} {\bibfield
  {journal} {\bibinfo  {journal} {Phys. Rev. D}\ }\textbf {\bibinfo {volume}
  {104}},\ \bibinfo {pages} {084007} (\bibinfo {year} {2021})},\ \Eprint
  {http://arxiv.org/abs/2006.10077} {arXiv:2006.10077 [gr-qc]} \BibitemShut
  {NoStop}%
\bibitem [{\citenamefont {Landau}\ and\ \citenamefont
  {Lifschits}(1975)}]{Landau:1975pou}%
  \BibitemOpen
  \bibfield  {author} {\bibinfo {author} {\bibfnamefont {L.~D.}\ \bibnamefont
  {Landau}}\ and\ \bibinfo {author} {\bibfnamefont {E.~M.}\ \bibnamefont
  {Lifschits}},\ }\href@noop {} {\emph {\bibinfo {title} {{The Classical Theory
  of Fields}}}},\ \bibinfo {series} {Course of Theoretical Physics}, Vol.\
  \bibinfo {volume} {Volume 2}\ (\bibinfo  {publisher} {Pergamon Press},\
  \bibinfo {address} {Oxford},\ \bibinfo {year} {1975})\BibitemShut {NoStop}%
\bibitem [{\citenamefont {{Schneider}}\ \emph {et~al.}(1992)\citenamefont
  {{Schneider}}, \citenamefont {{Ehlers}},\ and\ \citenamefont
  {{Falco}}}]{1992grle.book}%
  \BibitemOpen
  \bibfield  {author} {\bibinfo {author} {\bibfnamefont {P.}~\bibnamefont
  {{Schneider}}}, \bibinfo {author} {\bibfnamefont {J.}~\bibnamefont
  {{Ehlers}}}, \ and\ \bibinfo {author} {\bibfnamefont {E.~E.}\ \bibnamefont
  {{Falco}}},\ }\href {\doibase 10.1007/978-3-662-03758-4} {\emph {\bibinfo
  {title} {{Gravitational Lenses}}}}\ (\bibinfo  {publisher} {Springer},\
  \bibinfo {year} {1992})\BibitemShut {NoStop}%
\bibitem [{\citenamefont {Takahashi}(2004)}]{TakahashiD}%
  \BibitemOpen
  \bibfield  {author} {\bibinfo {author} {\bibfnamefont {R.}~\bibnamefont
  {Takahashi}},\ }\emph {\bibinfo {title} {{Wave Effects in the Gravitational
  Lensing of Gravitational Waves from Chirping Binaries}}},\ \href@noop {}
  {Ph.D. thesis},\ \bibinfo  {school} {Kyoto University} (\bibinfo {year}
  {2004}),\ \bibinfo {note}
  {\url{http://cosmo.phys.hirosaki-u.ac.jp/takahasi/dt.pdf}}\BibitemShut
  {NoStop}%
\bibitem [{\citenamefont {Bray}(1986)}]{Bray:1985ew}%
  \BibitemOpen
  \bibfield  {author} {\bibinfo {author} {\bibfnamefont {I.}~\bibnamefont
  {Bray}},\ }\href {\doibase 10.1103/PhysRevD.34.367} {\bibfield  {journal}
  {\bibinfo  {journal} {Phys. Rev. D}\ }\textbf {\bibinfo {volume} {34}},\
  \bibinfo {pages} {367} (\bibinfo {year} {1986})}\BibitemShut {NoStop}%
\bibitem [{\citenamefont {He}\ and\ \citenamefont {Wu}(2022)}]{He:2022sjf}%
  \BibitemOpen
  \bibfield  {author} {\bibinfo {author} {\bibfnamefont {J.-h.}\ \bibnamefont
  {He}}\ and\ \bibinfo {author} {\bibfnamefont {Z.}~\bibnamefont {Wu}},\ }\href
  {\doibase 10.1103/PhysRevD.106.124037} {\bibfield  {journal} {\bibinfo
  {journal} {Phys. Rev. D}\ }\textbf {\bibinfo {volume} {106}},\ \bibinfo
  {pages} {124037} (\bibinfo {year} {2022})},\ \Eprint
  {http://arxiv.org/abs/2208.01621} {arXiv:2208.01621 [gr-qc]} \BibitemShut
  {NoStop}%
\bibitem [{\citenamefont {Poisson}(2009)}]{Poisson:2009pwt}%
  \BibitemOpen
  \bibfield  {author} {\bibinfo {author} {\bibfnamefont {E.}~\bibnamefont
  {Poisson}},\ }\href {\doibase 10.1017/CBO9780511606601} {\emph {\bibinfo
  {title} {{A Relativist's Toolkit: The Mathematics of Black-Hole
  Mechanics}}}}\ (\bibinfo  {publisher} {Cambridge University Press},\ \bibinfo
  {year} {2009})\BibitemShut {NoStop}%
\bibitem [{\citenamefont {Isoyama}\ \emph {et~al.}(2013)\citenamefont
  {Isoyama}, \citenamefont {Fujita}, \citenamefont {Nakano}, \citenamefont
  {Sago},\ and\ \citenamefont {Tanaka}}]{Isoyama:2013yor}%
  \BibitemOpen
  \bibfield  {author} {\bibinfo {author} {\bibfnamefont {S.}~\bibnamefont
  {Isoyama}}, \bibinfo {author} {\bibfnamefont {R.}~\bibnamefont {Fujita}},
  \bibinfo {author} {\bibfnamefont {H.}~\bibnamefont {Nakano}}, \bibinfo
  {author} {\bibfnamefont {N.}~\bibnamefont {Sago}}, \ and\ \bibinfo {author}
  {\bibfnamefont {T.}~\bibnamefont {Tanaka}},\ }\href {\doibase
  10.1093/ptep/ptt034} {\bibfield  {journal} {\bibinfo  {journal} {PTEP}\
  }\textbf {\bibinfo {volume} {2013}},\ \bibinfo {pages} {063E01} (\bibinfo
  {year} {2013})},\ \Eprint {http://arxiv.org/abs/1302.4035} {arXiv:1302.4035
  [gr-qc]} \BibitemShut {NoStop}%
\bibitem [{\citenamefont {Wald}(1984)}]{Wald:1984rg}%
  \BibitemOpen
  \bibfield  {author} {\bibinfo {author} {\bibfnamefont {R.~M.}\ \bibnamefont
  {Wald}},\ }\href {\doibase 10.7208/chicago/9780226870373.001.0001} {\emph
  {\bibinfo {title} {{General Relativity}}}}\ (\bibinfo  {publisher} {Chicago
  Univ. Pr.},\ \bibinfo {address} {Chicago, USA},\ \bibinfo {year}
  {1984})\BibitemShut {NoStop}%
\bibitem [{\citenamefont {Hawking}\ and\ \citenamefont
  {Ellis}(2011)}]{Hawking:1973uf}%
  \BibitemOpen
  \bibfield  {author} {\bibinfo {author} {\bibfnamefont {S.~W.}\ \bibnamefont
  {Hawking}}\ and\ \bibinfo {author} {\bibfnamefont {G.~F.~R.}\ \bibnamefont
  {Ellis}},\ }\href {\doibase 10.1017/CBO9780511524646} {\emph {\bibinfo
  {title} {{The Large Scale Structure of Space-Time}}}},\ Cambridge Monographs
  on Mathematical Physics\ (\bibinfo  {publisher} {Cambridge University
  Press},\ \bibinfo {year} {2011})\BibitemShut {NoStop}%
\bibitem [{\citenamefont {Misner}\ \emph {et~al.}(1973)\citenamefont {Misner},
  \citenamefont {Thorne},\ and\ \citenamefont {Wheeler}}]{Misner:1973prb}%
  \BibitemOpen
  \bibfield  {author} {\bibinfo {author} {\bibfnamefont {C.~W.}\ \bibnamefont
  {Misner}}, \bibinfo {author} {\bibfnamefont {K.~S.}\ \bibnamefont {Thorne}},
  \ and\ \bibinfo {author} {\bibfnamefont {J.~A.}\ \bibnamefont {Wheeler}},\
  }\href@noop {} {\emph {\bibinfo {title} {{Gravitation}}}}\ (\bibinfo
  {publisher} {W. H. Freeman},\ \bibinfo {address} {San Francisco},\ \bibinfo
  {year} {1973})\BibitemShut {NoStop}%
\bibitem [{\citenamefont {Rybicki}\ and\ \citenamefont
  {Lightman}(2004)}]{Rybicki:2004hfl}%
  \BibitemOpen
  \bibfield  {author} {\bibinfo {author} {\bibfnamefont {G.~B.}\ \bibnamefont
  {Rybicki}}\ and\ \bibinfo {author} {\bibfnamefont {A.~P.}\ \bibnamefont
  {Lightman}},\ }\href {\doibase 10.1002/9783527618170} {\emph {\bibinfo
  {title} {{Radiative Processes in Astrophysics}}}}\ (\bibinfo  {publisher}
  {Wiley-VCH},\ \bibinfo {year} {2004})\BibitemShut {NoStop}%
\end{thebibliography}%

\end{document}